\documentclass[journal]{IEEEtran}
\usepackage{graphicx,epsfig,amsmath,amssymb,bm}
\usepackage{amssymb,balance}
\usepackage{amsmath}
\usepackage{amsthm} 
\usepackage{amssymb}
\usepackage{algorithmicx}
\usepackage{algpseudocode}
\usepackage{url}
\def\LB{\left(}         
\def\RB{\right)}        

\newfont{\bbb}{msbm10 scaled 500}

\newfont{\bb}{msbm10 scaled 1100}

\newcommand{\RR}{\mbox{\bb R}}

\newcommand{\bv}{{\bf b}}

\newcommand{\fv}{{\bf f}}

\newcommand{\kv}{{\bf k}}

\newcommand{\pv}{{\bf p}}

\newcommand{\xv}{{\bf x}}
\newcommand{\yv}{{\bf y}}
\newcommand{\zv}{{\bf z}}

\newcommand{\Bm}{{\bf B}}

\newcommand{\Dm}{{\bf D}}

\newcommand{\Id}{{\bf I}}

\newcommand{\Km}{{\bf K}}

\newcommand{\Mm}{{\bf M}}

\newcommand{\Om}{{\bf O}}

\newcommand{\Xm}{{\bf X}}

\newcommand{\deltav}{\hbox{\boldmath$\delta$}}

\newcommand{\epsilonv}{\hbox{\boldmath$\epsilon$}}

\newcommand{\thetav}{\hbox{\boldmath$\theta$}}

\newcommand{\omegav}{\hbox{\boldmath$\omega$}}

\newcommand{\varphiv}{{\bf \varphi}}

\renewcommand{\arg}{{\hbox{arg}}}

\renewcommand{\Re}{{\rm Re}}

\usepackage{bbm}
\usepackage{subcaption}
\usepackage{tabularx}
\usepackage{multirow}
\usepackage{verbatim}
\usepackage[ruled,linesnumbered]{algorithm2e}

\usepackage{color}
\newcounter{savefig}

 \newcommand{\req}[1]{(\ref{#1})}

 \def\b0{\mbox{\boldmath $0$}}
 \def\bbR{\mbox{$\boldsymbol{\mathbb R}$}}
 \def\bbC{\mbox{$\boldsymbol{\mathbb C}$}}

 \def\balpha{\mbox{\boldmath $\alpha$}}

 \def\bC{\mbox{\bf C}}

 \def\bG{\mbox{\bf G}}
 \def\bH{\mbox{\bf H}}

 \def\bM{\mbox{\bf M}}
 
 \def\bO{\mbox{\bf O}}

 \def\bu{\mbox{\bf u}}
 \def\bv{\mbox{\bf v}}
 
 \def\by{\mbox{\bf y}}
 \def\bz{\mbox{\bf z}}
 \def\bp{\mbox{\bf p}}
 \def\bff{\mbox{\bf f}}

 \def\bcalA{\mbox{\boldmath $\mathcal A$}}
 \def\bcalB{\mbox{\boldmath $\mathcal B$}}
 \def\bcalC{\mbox{\boldmath $\mathcal C$}}

 \def\bcalG{\mbox{\boldmath $\mathcal G$}}

 \def\bcalM{\mbox{\boldmath $\mathcal M$}}

\def\argmax{\operatornamewithlimits{arg\,max}}
\def\argmin{\operatornamewithlimits{arg\,min}}

\newtheorem{proposition}{Proposition}

\newcommand{\beqa}{\begin{eqnarray}}
\newcommand{\eeqa}{\end{eqnarray}}
\newcommand{\dsp}{\displaystyle}

\newcommand{\derip}[1]{\ensuremath{{\frac {\partial #1} {\partial \alpha_m} } } }

\ifCLASSINFOpdf

\else

\fi

\begin{document}

\title{Analysis of IoT-Based Load Altering Attacks Against Power Grids Using the Theory of Second-Order Dynamical Systems}
\author{Subhash~Lakshminarayana~\IEEEmembership{Senior Member, IEEE}, Sondipon Adhikari, and Carsten Maple \vspace{-0.35in}
\thanks{S. Lakshminarayana is with the School of Engineering, University of Warwick, Coventry, UK, CV47AL (email: subhash.lakshminarayana@warwick.ac.uk). S. Adhikari is
with College of Engineering, Swansea University, UK  (email: s.adhikari@swansea.ac.uk).
C. Maple is with the Warwick Manufacturing Group, University of Warwick, UK, CV47AL (email: cm@warwick.ac.uk).
}}

\maketitle

\begin{abstract}
Recent research has shown that large-scale Internet of Things (IoT)-based load altering attacks can have a serious impact on power grid operations such as causing unsafe frequency excursions and destabilizing the grid's control loops. In this work, we present an analytical framework to investigate the impact of IoT-based static/dynamic load altering attacks (S/DLAAs) on the power grid's dynamic response. Existing work on this topic has mainly relied on numerical simulations and, to date, there is no analytical framework to identify the victim nodes from which that attacker can launch the most impactful attacks. To address these shortcomings, we use results from second-order dynamical systems to analyze the power grid frequency control loop under S/DLAAs. We use parametric sensitivity of the system's eigensolutions to identify victim nodes that correspond to the \emph{least-effort} destabilizing DLAAs. Further, to analyze the SLAAs, we present closed-form expression for the system's frequency response in terms of the attacker's inputs, helping us characterize the minimum load change required to cause unsafe frequency excursions. Using these results, we formulate the defense against S/DLAAs as a linear programming problem in which we determine the minimum amount of load that needs to be secured at the victim nodes to ensure system safety/stability. Extensive simulations conducted using benchmark IEEE-bus systems validate the accuracy and efficacy of our approach.
\end{abstract}


\IEEEpeerreviewmaketitle

\section{Introduction}
The electric grid is undergoing a fundamental transformation from a centralized, producer-controlled network to one that integrates distributed players in its operations. Programs such as demand response seek the active involvement of end-users in reducing the grid's peak demand. Moreover, there is also a growing integration of Internet-of-Things (IoT) enabled devices at the consumer side, such as Wi-Fi-enabled air conditioners and residential battery energy storage systems \cite{Bosch}, which can be remotely controlled using personal computers or mobile phones such as smartphones, PC/tablets. These intelligent devices provide convenience, efficiency and monitoring capabilities, enabling consumers to better manage their usage.

However, IoT-enabled consumer appliances are often poorly engineered from a security point of view \cite{Fernandes2016HomeApp, maple2017security}. As such, they may become convenient entry points for malicious parties to gain access to the system and disrupt important grid operations by abruptly changing the demand. Cyber attacks targeting bulk power grid operations and state estimation problems have received significant attention \cite{Liu2009, RenLoadRedis2011, LaksheEnergy2017, LakshDataDrive2020, LakshCCPA2019}. In contrast, research on cyber attacks that target the end-user consumer devices is relatively new. Although internet-based load altering attacks were first introduced in \cite{HamedLAA2011}, which identified various load devices that are vulnerable to such attacks and proposed defense strategies, it was only recently that large-scale load altering attacks were studied considering a IoT-Botnet type attack \cite{Dabrowski2017,  Dvorking2017,  Soltan2018}. These works showed sudden and abrupt manipulation of the power grid demand due to such attacks can increase the grid's operational cost, and in some cases, cause unsafe frequency excursions. While power grid protection mechanisms such as under frequency load shedding (UFLS) can prevent large-scale blackouts, nevertheless, load-altering attacks remain capable of causing a partition in bulk power systems and/or a controlled load shedding event \cite{HuangUSENIX2019}. The aforementioned works are representatives of the so-called \emph{static load altering attacks} (SLAAs), which involves a one-time manipulation of the demand.

More severe attacks are the so-called \emph{dynamic load altering attacks} (DLAAs), in which the attacker changes the amount of compromised load over time to follow a certain trajectory \cite{AminiLAA2018, AminiIdentification2019}. In contrast to SLAAs, DLAAs require the attacker to monitor certain power grid signals (e.g., frequency) and alter the load in response to the fluctuations of the signal. This is feasible due to the availability of inexpensive commercial sensors to monitor the grid frequency (e.g., see \cite{FreqMes}), and they can be installed at any power outlet of the grid. These devices are already installed in existing frequency-sensitive loads that participate in the grid frequency regulation \cite{ZhaoFreq2013}. While DLAAs require enhanced capabilities on the part of the attacker, they can have a much more severe impact on grid operations than SLAAs, such as destabilizing the power grid control loops \cite{AminiLAA2018}, leading to generator trips and cascading failures.



A major shortcoming of existing work on load altering attacks is that they either adopt a simulation-based approach  (e.g., \cite{Soltan2018}  to assess the impact of SLAAs) or employ methods such as root locus analysis (e.g., \cite{AminiLAA2018} to assess the impact of DLAAs). However, these approaches can be computationally expensive as they require exhaustive simulations or eigenvalue computations under all possible combinations of nodes that could be targeted by the attacker (in a coordinated multi-point attack). They do not provide any physical insights into the system under DLAAs and SLAAs. Moreover, to date, there is no analytical method to identify the nodes that are most vulnerable to DLAAs and SLAAs.  Amini et. al. \cite{AminiLAA2018}  also propose a defense against DLAAs based on securing a portion of the vulnerable loads. However, finding the locations and the amount of the loads which must be secured requires solving a non-convex pole placement optimization problem that is computationally complex. A concurrent work \cite{DLAAStorage2020} presents an alternative defense by the use of energy storage systems to compensate for the destabilizing effects of DLAAs. However, the design requires further research on tuning the control parameters to ensure system stability under DLAAs.

To address these shortcomings, in this work, we present an analytical and low-complexity approach to assess the system's vulnerabilities and identify the victim nodes that correspond to the ``least-effort" DLAAs that will destabilize the system or SLAAs that will cause unsafe frequency excursions. Here ``least effort" is in terms of the amount vulnerable load that needs to be compromised at the victim buses to achieve the aforementioned objectives. As in prior work on this topic \cite{AminiLAA2018, AminiIdentification2019, AcharyaPHEV2020}, we use linear swing equations in our analysis. Our approach is based on the theory of second-order dynamical systems \cite{book1b}.

We make two important contributions. First, to analyze DLAAs, we compute the system's \emph{parametric eigenvalue sensitivities}. The sensitivity factors are a linear approximation that predict how much the system's eigenvalues change due to an incremental change in the attack parameters. The sensitivities can then be used to predict the attack impact on the system's stability. A major advantage of this approach is that the parametric sensitivity factors need to be computed considering single-point attacks only (i.e., considering DLAA at only one node of the grid at a time). Since the sensitivities are a linear approximation, the eigenvalues of the system under a coordinated multi-point attack can be approximated using the sum of eigenvalue sensitivities of multiple single-point attacks. Thus, the impact of multi-point attacks can be predicted using results from the single-point attacks.
Moreover, the computation of the parametric sensitivity factors itself is computationally cheap. Using the sensitivity approach, we propose a defense strategy against DLAAs in which we compute the least-amount of load that needs to be secured at each of the victim nodes to ensure  system stability. The defense problem requires solving a simple linear programming problem, which is also computationally cheap.


Second, to analyze SLAAs, we present a closed-form expression of the system's dynamic response due to a sudden change in the system load, in terms of its eigensolutions. Using these expressions, we can compute the maximum fluctuation in the system's frequency response due to a unit change in the load at a particular victim node, which helps us identify the victim node corresponding to the least-effort SLAAs. The closed-form expression also enables us to formulate the defense against SLAAs as a linear programming problem, in which we compute the least-amount of load that needs to be secured at each of the victim nodes to ensure no unsafe frequency excursions due to SLAAs.

Our results show that the eigenvalues obtained by the parametric sensitivity approach can accurately predict the true eigenvalues of the system under DLAAs over a wide range of attacker's parameters. Moreover, they also accurately characterize the nodes corresponding to the least-effort DLAAs. Our results also provide closed-form expressions to characterize the minimum values of attack control parameters that will destabilize the system (for DLAAs) and minimum change in the system load that will cause unsafe frequency excursions (for SLAAs) in terms of the system's eigensolutions. Further, the proposed defense can efficiently secure the system against destabilizing DLAAs or SLAAs.


To the best of our knowledge, this work is the first to apply results from the theory of second-order dynamical systems to analyze load altering attacks against power grids. The theory has been provably applied extensively in vibration problems in civil, mechanical, and aerospace engineering \cite{mei97}. While eigenvalue sensitivities have been applied in the past in power systems research for planning and analysis purposes (see e.g., \cite{Smed1993, NamEigSens2000}), they have not been utilized in a power grid security context. In particular, the application of parametric sensitivity analysis of second-order systems to analyze DLAAs and SLAAs is novel; this is one of the important contributions of our work.

The rest of the paper is organized as follows. We present the system model in Sec.~\ref{sec:Prelim}. We provide a brief overview of the theory of second-order systems in Sec.~\ref{sec:Sec_Order_sys}. Using this theory, we analyze DLAAs and SLAAs in Sec.~\ref{sec:DLAA} and Sec.~\ref{sec:SLAA} respectively. We present the simulation results in Sec.~\ref{sec:Results} and conclude in Sec.~\ref{sec:Conc}.

{\bf Notations:} We use bold font lower case and upper case to denote vectors and matrices respectively. We denote the $i^\text{th}$ entry of vector $\xv$ by $x_i$ and the $(i,j)^{\text{th}}$ entry of a matrix $\Xm$ by $\Xm_{i,j}.$ The real and imaginary parts of a complex number $X$ are denoted by $\text{Re}(X)$ and $\text{Im}(X)$ respectively. We use $\bO$  to denote a matrix of all zeros, $\b0$ to denote a vector of zeros, and ${\bf I}$ to denote the identity matrix. We use $[\xv;\yv]$ to denote the concatenation of vectors $\xv$ and $\yv$.

\section{System Model}
\label{sec:Prelim}
{\bf Power Grid Model:} We consider a power grid consisting of a set of $\mathcal{N} = \{1,\dots,N \}$ buses. The set of buses are divided into 
generator buses $\mathcal{G} = \{g_1,\dots,g_{N_G} \}$ and load buses $\mathcal{L} = \{l_1,\dots,l_{N_L} \}$, where $N_G$ and $N_L$ represent the number of generator and load buses respectively and $\mathcal{N} = \mathcal{G} \cup \mathcal{L}$. Here in, $g_i$ and $l_i$ represent the index of the $i^{\text{th}}$ generator and load bus respectively. We let $\pv^L \in \RR^{N_L}$ denote the vector of demands at the load buses $\mathcal{L}$. 
The linearized version of the power grid dynamic model is given by the differential equations \cite{kundur1994}:
\begin{align}
& \begin{bmatrix} 
\Id & {\bf O} & {\bf O} & {\bf O} \\
{\bf O}  & \Id  & {\bf O} & {\bf O} \\
{\bf O} & {\bf O} & -\Mm & {\bf O} \\
{\bf O} & {\bf O} & {\bf O} & {\bf O}
\end{bmatrix}
\begin{bmatrix} 
\dot{\deltav} \\
\dot{\thetav} \\
\dot{\omegav} \\
\dot{\varphiv}
\end{bmatrix} = \begin{bmatrix} 
{\bf 0} \\
{\bf 0} \\
{\bf 0} \\
\pv^{L} 
\end{bmatrix} + \nonumber \\
& \begin{bmatrix} 
{\bf O} & {\bf O} & \Id & {\bf O} \\
{\bf O} & \Id  & {\bf O}  & {\bf O} \\
\Km^I + \Bm^{GG} & \Bm^{GL} & \Km^P + \Dm^G  & {\bf O} \\
\Bm^{LG} & \Bm^{LL} & {\bf O} &  \Dm^{L}
\end{bmatrix}
\begin{bmatrix} 
{\deltav} \\
{\thetav}   \\
{\omegav} \\
\varphiv
\end{bmatrix}, \label{eqn:dyn_mtx}
\end{align}
where $\deltav, \omegav \in \RR^{N_G}$ comprise the phase angles and rotor frequency deviations at the generator buses respectively, 
$\thetav, \varphiv \in \RR^{N_L}$ comprise the phase angles and the frequency deviations of the load buses respectively. $\Mm,\Dm^G \in \RR^{N_G \times N_G}$ and $\Dm^L \in \RR^{N_L \times N_L}$ are diagonal matrices with their diagonal entries given by the generator inertia and generator damping coefficients and load damping coefficients respectively. 
$\Km^I,\Km^P \in \RR^{N_G \times N_G}$ are diagonal matrices with their diagonal entries given by the integral and proportional control coefficients of the generators respectively. Matrices $\Bm^{GG} \in \RR^{N_G \times N_G}, \Bm^{LL} \in \RR^{N_L \times N_L}, \Bm^{GL} \in \RR^{N_G \times N_L}$ are sub-matrices of the admittance matrix, derived as
$\Bm_{bus} = \begin{bmatrix} \Bm^{GG}  & \Bm^{GL} \\ 
\Bm^{LG} & \Bm^{LL}
\end{bmatrix}.$
We denote $\omega_{\text{nom}}$ as the grid's nominal frequency, e.g., $50~$Hz in Europe or  $60~$Hz in North America. For safe operations, the frequency must be maintained within the safety limits. We denote $\omega_{\text{max}}$ as the maximum permissible frequency deviation for system safety. Thus, $| \omega_{\text{nom}} - \omega_i| \leq \omega_{\text{max}}, \forall i \in \mathcal{G}.$ We note that in steady state, $\dot{\omega}_i = 0, \forall i \in \mathcal{G}.$


{\bf Load Altering Attacks:}
Under IoT-based load-altering attacks, the attacker manipulates the system load by synchronously switching on or off a large number of high-wattage devices. Assume that the demand at the load buses consists of two components
$\pv^L = \pv^{LS} + \pv^{LV},$ where $ \pv^{LS}$ denotes the secure part of the load (i.e., load that cannot be altered) and  $ \pv^{LV}$ denotes the vulnerable part of the load. 
We denote the set of victim nodes by $\mathcal{V} (\subseteq \mathcal{L}),$ and $N_v = |\mathcal{V} |,$ which are the subset of load buses at which the attacker can manipulate the load.  
The system load under load-altering attacks, which we denote by $\pv^L_a,$ is given by
\begin{align}
\pv^L_a = {\pv}^{LS}+\epsilonv^L - \Km^{LG} \omegav  - \Km^{LL} \varphiv. \label{eqn:load_attack}
\end{align}
Herein, $\epsilonv^L$ is a step-change in the load introduced by the attacker. Note $\epsilon^L_i = 0$ if $i \notin \mathcal{V}.$ The components $\Km^{LG} \omegav$ and  $\Km^{LL} \varphiv $ are time-varying load altering attacks that follows the frequency fluctuations of the generator buses and load buses respectively (note that these are a series of load alterations whose magnitude is varying with time). We assume that the attacker can monitor the frequency at a subset of the generator buses $\mathcal{S}_G (\subseteq \mathcal{N}_G),$ and/or a subset of load buses $\mathcal{S}_L (\subseteq \mathcal{N}_L),$ by accessing the frequency measuring devices at these nodes. We denote $\mathcal{S} = \mathcal{S}_G \cup \mathcal{S}_L.$ 
$\Km^{LG} \in \RR^{N_L \times N_G}$ and $\Km^{LL} \in \RR^{N_L \times N_L}$ denote matrices consisting of attack controller gain values, where the elements $\Km^{LG}_{i,j}$ are the gains corresponding to the attack at load bus $i \in \mathcal{N}_L$ that follows the frequency at generator bus $j \in \mathcal{N}_G.$ Similarly, $\Km^{LL}_{i,j}$ are the gains that follows the frequency at the load bus $j \in \mathcal{N}_L.$
Note that $\Km^{LG}_{i,j} = 0$ or $\Km^{LL}_{i,j} = 0$ either if $i \notin \mathcal{V}$ or  $j \notin \mathcal{S}.$ 
The power grid dynamics \eqref{eqn:dyn_mtx} with the load altering attack in \eqref{eqn:load_attack} becomes
\begin{align}
& \begin{bmatrix} 
\Id & {\bf O} & {\bf O} & {\bf O} \\
{\bf O}  & \Id  & {\bf O} & {\bf O} \\
{\bf O} & {\bf O} & -\Mm & {\bf O} \\
{\bf O} & {\bf O} & {\bf O} & {\bf O}
\end{bmatrix}
\begin{bmatrix} 
\dot{\deltav} \\
\dot{\thetav} \\
\dot{\omegav} \\
\dot{\varphiv}
\end{bmatrix} = \begin{bmatrix} 
{\bf 0} \\
{\bf 0} \\
{\bf 0} \\
\pv^{LS} + \epsilonv^L
\end{bmatrix} + \nonumber \\
& \begin{bmatrix} 
{\bf O} & {\bf O} & \Id & {\bf O} \\
{\bf O}& \Id  & {\bf O}  & {\bf O} \\
\Km^I + \Bm^{GG} & \Bm^{GL} & \Km^P + \Dm^G  & {\bf O} \\
\Bm^{LG} & \Bm^{LL} & -\Km^{LG} & - \Km^{LL} + \Dm^{L}
\end{bmatrix}
\begin{bmatrix} 
{\deltav} \\
{\thetav}   \\
{\omegav} \\
\varphiv
\end{bmatrix}. \label{eqn:dyn_mtx_attack}
\end{align}
The limits on the attack components are given as follows. $\epsilonv^L  \leq \pv^{LV},{K}^{L}_{v,s} \geq 0$ and 
\begin{align}
 \sum_{s \in \mathcal{S}} {K}^{L}_{v,s} \omega^{\max}_s   \leq (P^{LV}_v - \epsilon^L_v)/2 , \forall v \in \mathcal{V}. \label{eqn:Attack_lim}
\end{align}
The limit in \eqref{eqn:Attack_lim} can be explained as follows. First note that $ \sum_{s \in \mathcal{S}}  {K}^L_{v,s} \omega^{\max}_s $ is the maximum value of the load to be altered by the attacker at victim bus $v$ before the frequency at any sensor bus $s$ exceeds the  safety limit $\omega^{\max}_s.$ (Herein, we have used $\Km^L$ as a short-hand notation to denote either $\Km^{LG}$ or $\Km^{LL}.$)
This must be less than the amount of vulnerable load, i.e., $P^{LV}_v - \epsilon_v $ (after removing the step change). Finally, the factor of $2$ in the  denominator of the RHS represents the fact that the amount of load that can be compromised must allow for both over and under frequency fluctuations before the system frequency exceeds  $\omega^{\max}_s$ (see  \cite{AminiLAA2018}).



{\bf Problem Formulation:}
Within the framework of \eqref{eqn:load_attack}, we consider two types of load-altering attacks: (i) SLAAs, which consist of an abrupt one-time increase/decrease in power demand. In this case $\epsilonv^L \neq {\bf 0}$ and $\Km^{LG} = \Km^{LL} = \Om.$ As shown in \cite{Soltan2018}, if the value of $\epsilonv^L$ is large, this will result in unsafe frequency excursions. (ii) DLAAs, in which $\Km^{LG},  \Km^{LL} \neq \Om.$ Under DLAAs, the attacker can alter the eigenvalues of the system indirectly by changing the elements of the matrix $\Km^{LG}$ or $\Km^{LL}$. Thus, DLAAs can potentially destabilize the power grid frequency control loop \cite{AminiLAA2018}.

The objectives of this work are two-fold: (1)  identify the victim nodes that correspond to the least-effort SLAAs and DLAAs, i.e., buses from which an unsafe excursion or destabilizing attack can be launched by altering the least amount of load, and (2) find a low-computational defense strategy that computes the least amount of load to be secured at the victim nodes such that the attacker cannot launch a successful SLAAs or DLAAs. 
The results give us fundamental insights into identifying vulnerable nodes in the grid that are susceptible to DLAAs and SLAAs, and reinforce them to enhance the grid's resilience.

In the following section, we first provide a brief overview of general second-order systems and describe results from parametric sensitivities of its eigensolutions and dynamic response. Then, in Sections \ref{sec:DLAA} and \ref{sec:SLAA}, we apply these results to analyze DLAAs and SLAAs  respectively.



\section{Brief Review of General Second-Order Systems}
\label{sec:Sec_Order_sys}
Second-order matrix differential equations form the essential basis for the linear dynamic analysis of mechanical systems since their  introduction by Lord Rayleigh \cite{ray1877} in 1877. The standard second-order system is given by the following dynamic equation:
 \begin{align}
 \bcalM \ddot{\bu}(t) &+ \bcalC  \dot{\bu}(t)    + \bcalG  \bu(t)  = \bf{f}(t).  \label{eqn:second_order_dyn_gen}
 \end{align}
Here $\bu(t) \in \RR^{N}$ and ${\bf f}(t) \in \RR^{N}$ are the response vector and the forcing vector respectively. The system matrices in equation \req{eqn:second_order_dyn_gen}, namely  $\bcalM$, $\bcalC$ and $\bcalG \in \bbR^{N \times N}$, are the so-called inertia, damping and stiffness matrices. In general they are real and non-symmetric matrices. However, for many mechanical systems these matrices become symmetric matrices. In that case a simplified approach, known as the modal analysis \cite{mei97}, is available. Under certain conditions, a general non-symmetric system can be transformed into an equivalent symmetric system \cite{jp8}. For such symmetric linear systems, dynamic response in the frequency domain can be obtained efficiently \cite{li2014hybrid} using the eigenvalues and eigenvectors of the system.

\subsection{Eigenvalues and Eigenvectors of Second-Order System}
The eigenvalues and the eigenvectors are important descriptors of the system and they together determine the system's dynamic response. 
The \emph{right}
eigenvalue problem associated with the second-order system in\eqref{eqn:second_order_dyn_gen} can be
represented by the $\lambda-$matrix problem as
\begin{align*}
\lambda_j^2 \bcalM \bu_j + \lambda_j\bcalC \bu_j + \bcalG \bu_j = \b0,
\quad \forall \, j=1, \cdots ,{N}
\end{align*}
where $\lambda_j \in \bbC$ is the $j$-th latent root (eigenvalue) and ${\bf
	u}_j \in \bbC^N$ is the $j$-th  right latent vector (right eigenvector).
The \emph{left} eigenvalue problem can be represented by
\begin{align*}
\lambda_j^2 \bv_j^\top \bcalM + \lambda_j\bv_j^\top \bcalC + \bv_j^\top \bcalG = \b0^\top,
\quad \forall \, j=1, \cdots ,{N}
\end{align*}
where $\bv_j \in \bbC^N$ is the $j$-th left latent vector (left
eigenvector) and $(\bullet)^\top$ denotes the matrix transpose. When
${\bf M, C}$ and $\bG$ are general asymmetric matrices the right
and left eigenvectors can easily be obtained from the first-order
formulations, for example, the state-space method or
Duncan forms \cite{mei80}. Equation \req{eqn:second_order_dyn_gen} is transformed
into the first-order (Duncan) form as
\begin{equation}
\label{p4:eq0.5}
\bcalA \, \dot {\bf \mathfrak z} (t) +
\bcalB \, {\bf \mathfrak z} (t) =  {\bf \mathfrak f} (t) 
\end{equation}
where $\bcalA, \,  \bcalB \in \bbR^{{2N} \times 2{N}}$ are the system
matrices,  ${\bf \mathfrak f}(t) \in \bbR^{2N}$ is the forcing vector and ${\bf \mathfrak z}(t) \in \bbR^{2{N}}$ is the state vector
given by
\begin{align}
\label{p4:eq0.6}
\bcalA &=
\begin{bmatrix}
\bcalC & \bcalM \\
\bcalM &  \bO
\end{bmatrix},
\bcalB =
\begin{bmatrix}
\bcalG & \bO \\
\bO &  - \bcalM
\end{bmatrix}
\\
{\bf {\mathfrak f}} (t) &=
\begin{Bmatrix}
{\bf f}(t) \\
{\bf 0}
\end{Bmatrix},
{\bf {\mathfrak z}} (t)=
\begin{Bmatrix}
{\bf {\mathfrak u}}(t) \\
{\dot {\bf \mathfrak u}}(t)
\end{Bmatrix}.
\end{align}
Taking the
Laplace transform of equation \req{p4:eq0.5} we obtain
\begin{equation}
\label{eqn:lap_main}
s \bcalA {\bf \bar z} (s)+ \bcalB {\bf \bar z} (s) = 
  {\bf \bar f} (s)+ \bcalA \bz_0 
\end{equation}
Here, ${\bf \bar z}(s)$ is the Laplace transform of ${\bf \mathfrak z} (t)$ and $ {\bf \bar f} (s)$ is the Laplace transform of ${\bf \mathfrak f} (t)$ and the initial condition vector in the state-space  
${\bf \mathfrak z} (0) = \bz_0$. The vector $\bp(s) =  {\bf \bar f} (s)+ \bcalA \bz_0  $ is the effective state-space forcing function in the Laplace domain. 

The {\it right} and {\it left}  eigenvalue problem associated with
equation \req{p4:eq0.5} can be expressed as
\begin{align*}
\lambda_j \bcalA \bz_j + \bcalB \bz_j & = \b0,
\forall j=1,\cdots, 2{N}, \\
\lambda_j \by_j^\top \bcalA + \by_j^\top \bcalB & = \b0, \forall j=1,\cdots, 2{N},
\end{align*}
where $\lambda_j \in \bbC$ is the $j$-th eigenvalue and $\bz_j,  \by_j \in {\mathbb
	C}^{2{N_G}}$ is the $j$-th right/left eigenvector which is related to the $j$-th
right/left eigenvector of the second-order system as
$\bz_j = [
\bu_j ;
\lambda_j \bu_j]$ and $ \by_j = [
\bv_j ;
\lambda_j \bv_j].$

\subsection{Dynamic Response of the Second-Order System}
The eigenvalues along with the right and left eigenvectors of the first-order system can be used to obtain the dynamic response of the system in an efficient manner under general forcing and initial conditions. The transfer function matrix of the system in the Laplace domain can be expressed in terms of the eigensolutions (see for example \cite{jp7}) as 
$\bH(s) = \sum_{j=1}^{2{N_G}} {\frac { \bz_j \by_j^\top} { (s-\lambda_j)} }.$
Using this, the response vector can be obtained from equation \req{eqn:lap_main} as
\begin{equation}\label{eqn:dyn_resp_L}
{\bf \bar z} (s) = \bH(s) \bp(s) 
= \sum_{j=1}^{2{N_G}} {\frac { \by_j^\top \bp(s)} { (s-\lambda_j)} } \bz_j.
\end{equation}
This is the most general expression of the response vector as a function of the total forcing function $\bp(s)$ includes both the initial conditions and applied forcing. We consider a special case when the applied forcing function is a step function of the form ${\bf \mathfrak f} (t)  = U(t) \bff_0,$
where $\bff_0$ is a vector containing amplitudes of the forcing at different degrees of freedom and $U(\bullet)$ is a unit step function. 
Using this we obtain 
$\bp(s) =  {\frac{1}{s}} \bff_0  + \bcalA \bz_0.$
Substituting this expression of $\bp(s)$ in equation \req{eqn:dyn_resp_L} and taking the inverse Laplace transform of equation \req{eqn:dyn_resp_L}, we obtain
\begin{equation}\label{eqn:step_resp}
{\bf \mathfrak z} (t)  =
\sum_{j=1}^{2{N_G}}  a_j(t)   \bz_j,
\end{equation}
\begin{align}
\text{where} \ 
a_j(t) & =
\LB {\frac{e^{\lambda_jt}-1}{\lambda_j}} \RB
 \by_j^\top \bff_0
+ e^{\lambda_j t} \left (\by_j^\top \bcalA \bz_0 \right). \label{eqn:aj_def}
\end{align}
Equations \req{eqn:step_resp} and \eqref{eqn:aj_def} give the general closed-form expression of the response vector of non-symmetric second-order dynamic systems in terms of the eigensolutions.

\subsection{Parametric Sensitivity of the Eigensolutions}
A key interest in this paper is to quantify the change in the system characteristics and the response when elements of the system matrices change. To include all possible changes in the system matrices in a generic manner, we assume that the system matrices  $\bcalM$, $\bcalC$ and $\bcalG$ are functions of a parameter vector $\balpha = \left \{ \alpha_1,
\alpha_2, \cdots \alpha_m \right \}^\top \in {\mathbb R}^m$. As a result, the mass,
damping and stiffness matrices become functions of $\balpha$, that is $\bcalM \equiv \bcalM (\balpha), \bcalC \equiv \bcalC (\balpha)$ and $\bcalG \equiv \bcalG (\balpha)$. We consider these functions to be smooth, continuous and differentiable. There are several publication which discuss the parametric sensitivity of the eigensolutions of symmetric second-order systems (see for example \cite{book1b}). Below we follow the derivations in \cite{jp14} for non-symmetric second-order systems.

\subsubsection{Sensitivity of Eigenvalues}
We consider a generic element in the parameter vector $\alpha_m \in \balpha$. The sensitivity of the eigenvalue of a second-order system with respect to the parameter $\alpha_m$ is given by \cite{jp14}:
\begin{equation}
\frac{ \partial{\lambda_j}}{ \partial \alpha_m} 
 = -
\by_j^\top\left [ \lambda_j  \frac{ \partial \bcalA}{ \partial \alpha_m}  + \frac{ \partial \bcalB}{ \partial \alpha_m}  \right] \bz_j. \label{eqn:sens_general}
\end{equation}
 Note that the derivative of a given eigenvalue requires the knowledge of
only the corresponding eigenvalue and right and left eigenvectors
under consideration, and thus a complete solution of the
eigenproblem is not required. 

\subsubsection{Sensitivity of Eigenvectors}
The sensitivity of the eigenvector of a second-order system with respect to the parameter $\alpha_m$ is given by \cite{jp14}:
\begin{eqnarray}
\label{p4:eq2.3}
\frac{ \partial {\bz_j}} { \partial \alpha_m} = \sum_{l=1}^{2{N_G}} a^{(\alpha)}_{jl} \bz_l 
\ \text{and} \
  \frac{ \partial {\by_j}} { \partial \alpha_m}  = \sum_{l=1}^{2{N_G}} b^{(\alpha)}_{jl} \by_l.
\end{eqnarray}
Here $a^{(\alpha)}_{jl}$ and $b^{(\alpha)}_{jl}$,  $\forall \,
l=1, \cdots ,2N$ are sets of complex constants defined as 
\begin{align*}
a^{(\alpha)}_{jl}  =
- \by_l^\top \left [ \lambda_j  \frac{ \partial \bcalA}{ \partial \alpha_m}  + \frac{ \partial \bcalB}{ \partial \alpha_m}  \right] \bz_j,
 l=1,\cdots,2{N}; l \neq j,
\end{align*}
\begin{align*}
b^{(\alpha)}_{jl}  =
- \by_j^\top \left [ \lambda_j  \frac{ \partial \bcalA}{ \partial \alpha_m}  + \frac{ \partial \bcalB}{ \partial \alpha_m}  \right] \bz_l,
 l=1,\cdots,2{N}; l \neq j,
\end{align*}
\begin{align*}
\text{and} \ a^{(\alpha)}_{jj} = b^{(\alpha)}_{jj} =
- {\frac{1}{2}} \bv_j^\top
\left [ 2 \lambda_j  \derip{\bM} +  \derip{\bC} \right ]
\bu_j.
\end{align*}



\subsubsection{Sensitivity of the Step-Response}
Using the results above, we derive the sensitivity of the step response with respect to the change in the parameter $\alpha_m.$ The result is summarized in the following proposition.
\begin{proposition}
\label{prop:step_sens}
The parametric sensitivity of the step response with respect to $\alpha$ can be computed as
\begin{equation} 
\derip{{\bf \mathfrak z} (t)} = 
 \sum_{j=1}^{2{N_G}} \LB  \derip{a_j(t)}\bz_j +  a_j(t) \derip{\bz_j} \RB,
\end{equation}
where $\derip{a_j(t)} $ is a function of the eigenvalues and the eigenvectors and their derivatives. 
\end{proposition}
The expression of  $\derip{a_j(t)}$ and the derivation of Proposition~\ref{prop:step_sens} is  presented in a technical report \cite{TechReport}.


%
%

\section{Analysis of Dynamic Load Altering Attacks Based on Second-Order System Theory}
\label{sec:DLAA} 
We now employ the results presented in Section~\ref{sec:Sec_Order_sys} to analyze DLAAs. 
Using some straightforward manipulations, the power grid dynamic equations in  \eqref{eqn:dyn_mtx_attack} can be converted into the second-order system as
\begin{align}
& \underbrace{ \begin{bmatrix} 
 \Mm & {\bf O} \\
{\bf O} & {\bf O}
\end{bmatrix}}_{\bcalM}
\begin{bmatrix} 
\ddot{\omegav} \\
\ddot{\varphiv}
\end{bmatrix} 
+    
\underbrace{\begin{bmatrix} 
 \Km^P + \Dm^G & {\bf O} \\
-\Km^{LG} & \Km^{LL}
\end{bmatrix}}_{\bcalC}  \begin{bmatrix} 
\dot{\omegav} \\
\dot{\varphiv}
\end{bmatrix}
+ \nonumber \\
& \underbrace{\begin{bmatrix} 
 \Km^I + \Bm^{GG}  & \Bm^{GL}  \\
\Bm^{LG} & \Bm^{LL}
\end{bmatrix}}_{\bcalG}  
\begin{bmatrix} 
{\omegav} \\
{\varphiv}
\end{bmatrix}
= \underbrace{\begin{bmatrix} 
{\bf 0} \\
-(\pv^{LS} + \epsilonv^L)
\end{bmatrix}}_{\fv_0} . \label{eqn:dyn_mtx_attack_two}
\end{align}
Let $\{ \lambda_i \}^{2N }_{i = 1}$ denote the eigenvalues of the system without DLAAs (i.e., with $\Km^{LG} = \Km^{LL} = \Om$). Since the system is stable without attacks, we must have  $\Re(\lambda_i) < 0, i = 1,\dots,2N.$ 

In DLAAs, the attacker can indirectly control the system matrices by changing the elements of 
$\Km^{LG}$ or $\Km^{LL}.$ Let us denote the eigenvalues of the system with DLAAs by $\{ \nu_i (\Km^L)\}^{2 N}_{i = 1},$ where (with a slight abuse of notation) we have used $\Km^L$ as a short-hand notation to denote either $\Km^{LG}$ or $\Km^{LL}.$
The system will be rendered unstable if there exists at least one $\nu_i(\Km^L)$ such that  $\Re(\nu_i(\Km^L)) > 0.$ Thus, in order to understand the impact of DLAAs on  system stability, we must understand how the eigenvalues of the system change with respect an incremental change in the elements of the matrix $\Km^L.$ 

Our approach is to treat the elements of $\Km^{LG}$ and  $\Km^{LL}$ as parameters of the system matrices and use the parametric sensitivity results to analyze DLAAs. First note that the matrices ${\bcalM}$ and ${\bcalG}$ are independent of the elements of $\Km^{L}$, so they need not be consider them in the sensitivity analysis. The  matrix $\bcalC$ however is a smooth continuous, and differentiable function of the elements of $\Km^{L}$.
Hence, the method of sensitivity of second-order systems is directly applicable to the analysis of DLAAs. 
It can be shown that for the power grid model in \eqref{eqn:dyn_mtx_attack_two}, using \eqref{eqn:sens_general}, the parametric sensitivity of the eigenvalues with respect to elements of $\Km^L$ can be computed as
\begin{align}
\frac{\partial  \lambda_i}{\partial  K^L_{v,s}}   = -\lambda_i \yv^\top_i  \begin{bmatrix}
\bO & \bO \\
\frac{\partial  \bcalC }{\partial  K^{L}_{v,s}}  &  \bO
\end{bmatrix}   \zv_i, \label{eqn:sens_powergrid}
\end{align}
where, 
\begin{align*}
\frac{\partial  \bcalC }{\partial  K^{L}_{v,s}}  =
\begin{cases}
	 \begin{bmatrix} 
 {\bf O} & {\bf O} \\
-\Id_{v,s} & {\bf O}, 
\end{bmatrix},  & \text{if } \ s \in \mathcal{S}_G,    \\
	\begin{bmatrix} 
 {\bf O} & {\bf O} \\
{\bf O} & \Id_{v,s} , 
\end{bmatrix} & \text{if } \ s \in \mathcal{S}_L. 
\end{cases} 
\end{align*}
Here in, $\Id_{v,s} $ is a matrix whose $(v,s)^{\text{th}}$ entry is $1,$ and all other entries are zero. Note that $\frac{\partial  \lambda_i}{\partial  K^L_{v,s}} = 0$ if $v \notin \mathcal{V}$ and $s \notin \mathcal{S}.$ 
Using sensitivity analysis, the estimate $\widehat{\nu}_i ( \Km^L)$ of  $\nu_i (\Km^L)$ can be computed as:
\begin{align}
\widehat{\nu}_i ( \Km^L)= \lambda_i + \sum_{v \in \mathcal{V}} \sum_{s \in \mathcal{S}} \ \frac{\partial  \lambda_i}{\partial  K^L_{v,s}}  K^L_{v,s}, \  i = 1,\dots,2N. \label{eqn:Eigval_effect}
\end{align}



\subsection{Least-Effort DLAAs Using Parametric Sensitivity}
The node corresponding to the least-effort destabilizing single-point DLAA can be located using sensitivity analysis in the following manner. 
First note that if there exists at least one $\nu_i (\Km^L) > 0$ the system becomes unstable. Also, we assume that $\widehat{\nu}_i (\Km^L)$ closely approximates $\nu_i (\Km^L).$ Then, using \eqref{eqn:Eigval_effect} under single-point DLAA, it follows that a feedback gain greater than  $ \frac{-\lambda_i}{ \frac{\partial  \lambda_i}{\partial  K^L_{v,s}} }$
renders the eigenvalue $\widehat{\nu}_i ( \Km^L)$ to be positive. We denote the minimum value of the feedback gain at which the system becomes unstable by ${K}^{L^*}_{v,s}$ and its estimate by $\widehat{K}^{L^*}_{v,s}.$ Since only one eigenvalue is
required to be positive for the system to be unstable, it follows
that
\begin{align}
\widehat{K}^{L^*}_{v,s} = \min_{i = 1,\dots,2N_G} \frac{-\lambda_i}{ \frac{\partial  \lambda_i}{\partial  K^L_{v,s}} } \label{eqn:min_FBgain}
\end{align}
is the minimum value of feedback gain that makes the system unstable. Using \eqref{eqn:min_FBgain}, 
the node that corresponds to the least-effort destabilizing attack can be found as 
\begin{align}
\{v^*,s^*\} = \argmin_{v \in \mathcal{V}, s \in \mathcal{S}}  \widehat{K}^{L^*}_{v,s}.
\end{align}

A similar analysis can be performed for a coordinated multi-point DLAA. In particular, the set of feedback gain values that destabilize the system can be characterized as follows.
Let $\kv^L \in \mathbb{R}^{N_v N_s}$ denote a vector whose elements are given by  $K^L_{v,s}, v \in \mathcal{V}, s \in \mathcal{S}.$ If we  define a polyhedron $\mathcal{P}$ as
\begin{align*}
\mathcal{P} = \{ {\kv}^L |  \lambda_i + \sum_{v \in \mathcal{V}} \sum_{s \in \mathcal{S}} \ \frac{\partial  \lambda_i}{\partial  {K}^L_{v,s}}  {K}^L_{v,s} < 0, \  i = 1,\dots,2N_G  \},
\end{align*}
then all feedback gain vectors ${\kv}^L$ that lie outside $\mathcal{P}$ render the system unstable. 

Finally, note that the system dynamics under DLAAs can also be evaluated by using results from the parametric sensitivity of the step response presented in Proposition~\ref{prop:step_sens}.

\subsection{Defending Against DLAAs Based on Parametric Sensitivity Results}
Next, we illustrate the utility of the parametric sensitivity approach to defend against DLAAs. We adopt a similar approach to that of  \cite{AminiLAA2018}. The main idea to find the minimum amount of load that must be protected to ensure system stability in the face of DLAAs. In practical terms, protecting the load implies enhancing security measures such as enabling encryption at a device level or in the communication links. Minimizing the total amount of protected load will in turn minimize the cost of deploying such security measures. 
The defense problem can be formulated as a linear program (LP) as follows: 
\beqa
 & \dsp \min_{P^{LP}_v, {K}^L_{v,s}} &  \sum_{v \in \mathcal{V}} P^{LP}_v \label{eqn:Defense_DLAA}
    \\ 
& s.t. & 0 \leq P^{LP}_v \leq P^{LV}_v, \ \forall v \in \mathcal{V}, \nonumber
 \\
& &  \lambda^r_i +  \sum_{v \in \mathcal{V}} \sum_{s \in \mathcal{S}} \LB \frac{\partial  \lambda_i}{\partial  K^L_{v,s}} \RB^r {K}^L_{v,s} < 0, \  \forall \{ \lambda_i \}^{2 N_G}_{i = 1}, \nonumber \\
& &  \sum_s  {K}^L_{v,s} \omega^{\max}_s = (P^{LV}_v - P^{LP}_v)/2, \forall v \in \mathcal{V}, \nonumber
\eeqa  
where $X^r$ represents $\Re(X).$
In \eqref{eqn:Defense_DLAA}, $P^{LP}_v$ denotes the amount of load (from the vulnerable portion of the load) that must be protected at victim node $v \in \mathcal{V}.$ Naturally, this must be less than the total vulnerable load (first constraint of \eqref{eqn:Defense_DLAA}). The second constraint of \eqref{eqn:Defense_DLAA} ensures that the eigenvalues are negative, and hence, the system cannot be made unstable by the DLAA. The final constraint represents the limit on the attack controller's gain, which follows from \eqref{eqn:Attack_lim}. In this constraint, we use equality to ensure that the system remains stable even if the attacker uses the maximum permissible value of the attack controller gain. This is because the defender does not have prior knowledge of the actual parameters that the attacker intends to use. Without such knowledge, the defender must design a defense that is capable of ensuring system stability against all possible attack parameters.

 

We note that although the main idea behind the defense (i.e., protecting the vulnerable load) is similar to \cite{AminiLAA2018}, a key advantage of our approach is that it only requires solving an LP rather than solving a non-convex pole placement optimization problem. LPs can be solved exactly and efficiently, demonstrating the effectiveness of the proposed approach of analyzing DLAAs using the parametric sensitivity analysis of the eigensolutions.

{\bf Computational Complexity:} The main advantage of the parametric sensitivity approach is the reduction in computational complexity. 
We note that the sensitivity parameters only need to be computed for single-point attacks only (i.e., one victim node at a time). This amounts to computing $ N_v N_s$ sensitivity factors for each eigenvalue of the system. Moreover, the computations  \eqref{eqn:sens_powergrid} are cheap, since $\lambda_i, \yv_i$ and $\zv_i$ need to only be computed once. Only the factor $\Bm^M \Id_{v,s}$ must be recomputed for every combination of victim and sensor nodes. 
For a coordinated multipoint attack, the net effect of attacking multiple nodes can be computed using the superposition principle as in \eqref{eqn:Eigval_effect}. In contrast, directly assessing the impact of DLAAs would require recomputing the eigenvalues for each combination of victim/sensor nodes (there are $2^{N_v N_s} $ such combinations) and each value of the feedback gain. This method soon becomes computationally infeasible.

\section{Analysis of Static Load Altering Attacks}
\label{sec:SLAA} 
In this section, we analyze SLAAs using results from the theory of second-order systems.
From \eqref{eqn:dyn_mtx_attack_two}, note that under SLAAs, the attacker cannot modify the system matrices. As such, it is not possible to change the eigenvalues of the system.  Thus the parametric sensitivity analysis of eigensolutions cannot be used to assess the attack impact directly.
 
Although SLAAs cannot destabilize the system,  a sudden and abrupt change in the system load can result in unsafe frequency excursions \cite{Soltan2018}. It is thus of interest to identify nodes that correspond to the least-effort SLAAs (in terms of the amount of altered load) and defend the system against such attacks. To this end, we use the analytical expression for the step response presented in \eqref{eqn:step_resp} and \eqref{eqn:aj_def}. Without loss of generality, we assume initial conditions $\zv_0 = {\bf 0}.$ Thus, the step response in the time domain is given by
\begin{equation}
{\bf \mathfrak z} (t)  =
\LB \sum_{j=1}^{2N_G}  {\frac{e^{\lambda_jt}-1}{\lambda_j}} \RB
 (\by_j^\top \fv_{ 0})   \bz_j. \label{eqn:step_resp_grid}
\end{equation}
In the above equation, note that ${\bf \mathfrak z} (t) = [\deltav(t);\omegav(t)].$
Using \eqref{eqn:step_resp_grid}, we can express the power grid response to a change in the system load $\epsilonv^L.$ First note from \eqref{eqn:dyn_mtx_attack_two}, the forcing function $\fv_0$ and the change in the load $\epsilonv^L$ are related as $\fv_0 = [ {\bf 0}; - (\pv^{LS} + \epsilonv^L)]$.
Using this in \eqref{eqn:step_resp_grid}, and rearranging, we obtain,
\begin{align}
{\bf \mathfrak z} (t)  = \sum^{N_L}_{i = 1} \epsilon^L_{l_i} \sum^{2N_G}_{j = 1} \LB {\frac{e^{\lambda_jt}-1}{\lambda_j}} \RB k_{ji} \zv_j, \label{eqn:response_load}
\end{align}
where $\kv_j $ is a row vector given by $\kv_j  = \yv_j^\top   \in \RR^{1 \times N_L}$ and $\kv_{ji}$ is the $i^{\text{th}}$ element of $\kv_j .$ For convenience, let us denote 
\begin{align}
\fv_i(t) = \sum^{2N_G}_{j = 1} \LB {\frac{e^{\lambda_jt}-1}{\lambda_j}} \RB k_{ji} \zv_j.
\end{align}
Note that $\fv_i(t) = [f_{i,1}(t), \dots,f_{i,2N_G}(t)]^\top$ at each time $t$ is a $2N_G$ dimensional vector, where each element of $\fv_i(t)$ corresponds to the fluctuation of the components of  ${\bf \mathfrak z} (t).$


Equation \eqref{eqn:response_load} gives us a closed-form expression of the system's response in terms of the change in the load. A salient observation is that the response is a linear function of the load perturbation $\epsilonv^L.$ Assume we are interested in the fluctuation of the frequency at the $n^{\text{th}}$ generator bus. Let us define 
\begin{align}
l_{i^*,n} = \argmax_{i = 1,\dots,N_L} f_{i,n} (t^*_{i,n}), n = N_G+1,\dots,2N_G, \label{eqn:argmaxf}
\end{align}
where $t^*_{i,n} = \argmax_t f_{i,n}(t).$
Since  \eqref{eqn:response_load} is a linear function of the system load, under a single-point SLAA, the node $l_{i^*_n}$ is the node that corresponds to least-effort attack.
In \eqref{eqn:argmaxf}, $t^*_{i,n}$ can be found by simply taking the derivative of $f_{i,n}(t)$ with respect to time and finding the time at which the derivative function is zero. Note that there may be multiple times at which the derivative function becomes zero. For a stable system, we must have $\Re(\lambda_i) < 0.$ Thus, each function $f_{i,n}(t)$ is a decaying function of time (note the component $e^{\Re(\lambda_j)})$ in the numerator of \eqref{eqn:response_load}). It thus follows that the peak fluctuation of $f_{i,n}(t)$ occurs at time $t$ at which the derivative function becomes zero for the first time.

The function $f_{i,n}(t)$ represents the fluctuation of the $n^\text{th}$ frequency component for a per unit change in the system load. Thus, the minimum load change at load bus $i$ under SLAA that can cause an unsafe frequency excursion in the $n^\text{th}$ generator bus frequency can be computed as
\begin{align}
\epsilon^L_{l_{i},n} = \frac{\omega^{max}_n}{f_{i,n} (t^*_{i,n})}, i = 1,\dots,N_L. \label{eqn:least_load}
\end{align} 
As in the case of DLAAs, the closed-form expression of the step response in \eqref{eqn:response_load} can also be used to formulate the defense optimization problem against SLAAs as follows:
\beqa
 & \dsp \min &  \sum_{v \in \mathcal{V}} P^{LP}_v \label{eqn:Defense_SLAA}
    \\ 
& s.t. & 0 \leq P^{LP}_v \leq P^{LV}_v, \ \forall v \in \mathcal{V}, \nonumber
 \\
& &   \Big{|}  \sum^{N_L}_{i = 1} \epsilon^L_i \sum^{2N_G}_{j = 1} \LB {\frac{e^{\lambda_jt^*_{i,n}}-1}{\lambda_j}} \RB k_{ji} \zv_{jn} \Big{|} \leq \omega^{\max}_n \nonumber  \\ & & \qquad \qquad \qquad \  n = 1,\dots,2N_G, \nonumber \\
& & \epsilon^L_v = P^{LV}_v - P^{LP}_v, \forall v \in \mathcal{V}.  \nonumber
\eeqa  
In \eqref{eqn:Defense_SLAA}, the objective function and the first constraint equation is similar to \eqref{eqn:Defense_DLAA}. The second constraint represents the fact that the peak of the system's response due to the SLAA must not exceed the safety limit. The last constraint ensures that the compromised load does not exceed the vulnerable portion of the load after protection. As in \eqref{eqn:Defense_DLAA}, we consider equality constraint to ensure no unsafe frequency excursions even if the attacker alters the maximum permissible load. 

Once again, we note that optimization \eqref{eqn:Defense_SLAA} is a linear programming problem, which can be solved exactly and efficiently. This again illustrates the merit of the proposed technique. 
Finally, note that combined defense against DLAAs and SLAAs can be solved in a straightforward manner by combining the constraints of \eqref{eqn:Defense_SLAA} and \eqref{eqn:Defense_DLAA}. We omit the details due to the lack of space.

\section{Simulation Results}
\label{sec:Results}
In this section, we present simulation results to illustrate the effectiveness of the proposed approach. All simulations are conducted using MATLAB and  the power grid topological data obtained from the MATPOWER simulator. We use IEEE-6, 14, and 39 bus systems to illustrate our results. The power grid dynamics are obtained by solving the differential equations \eqref{eqn:dyn_mtx} in MATLAB.

\begin{figure}[!t]
\centering
\begin{subfigure}{0.45\textwidth}
\includegraphics[width=1\textwidth]{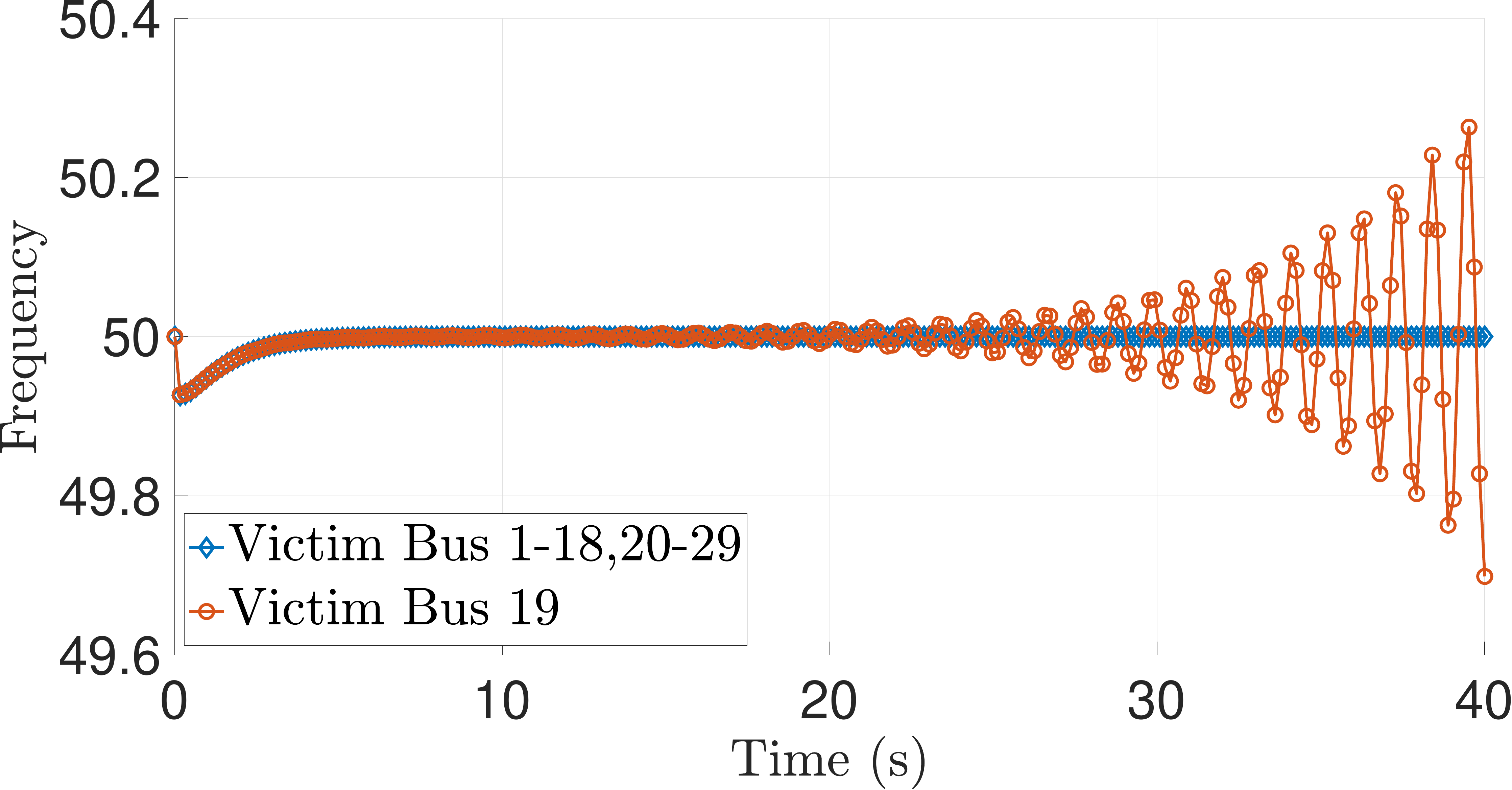}
\caption{}
\end{subfigure}
~
\begin{subfigure}{0.45\textwidth}
\includegraphics[width=1\textwidth]{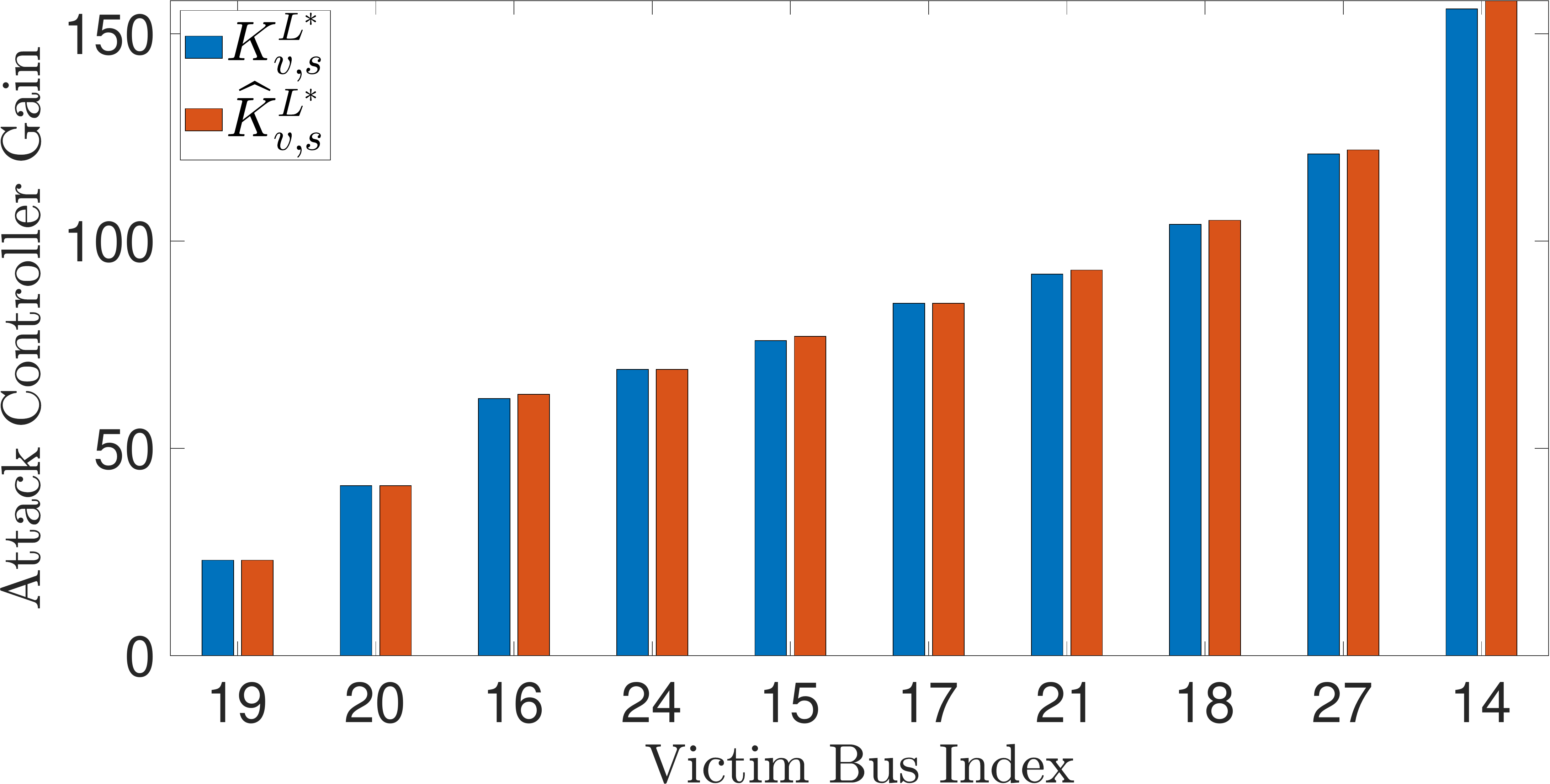}
\caption{}
\end{subfigure}
\caption{(a) Frequency dynamics under single-point DLAAs for different victim buses in the IEEE-39 bus system, $K^L_{v,s} = 25$ with $s = 33$ (sensing bus). Maximum compromised load = $2~$p.u. (b) Values of $K^{L^*}_{v,s}$ and $\widehat{K}^{L^*}_{v,s}$ for different victim buses.}
\label{fig:Dyn_grid}
\vspace{-0.1 cm}
\end{figure}

{\bf Single-Point DLAAs:} First we consider single-point DLAAs using the IEEE 39-bus system. We set bus~33 as the sensing bus, and inject DLAAs at the load buses~1-29 (one bus at a time). At victim bus $v \in \mathcal{V},$ we set the attack controller gain $K^L(v,33) = 25,$ ($v \in \{ 1-29 \}$), which corresponds to a maximum compromised load of $2 K^L_{v,s} \omega_{\max} = 2 \times 25 \times 2/50 = 2~\text{p.u.}$
We plot the frequency dynamics under DLAAs in Fig.~\ref{fig:Dyn_grid}~(a). 
For clarity, we only plot the frequency dynamics of one of the first generator buses, i.e., bus~30. We observe that the DLAA at victim bus~$19$ is able to destabilize the system, whereas the DLAA rest of the victim buses do not. Thus, bus~$19$ corresponds to the least-effort destabilizing attack. We also plot the values of $K^{L^*}_{v,s}$  and $\widehat{K}^{L^*}_{v,s}$ for $10$ different victim nodes in Fig.~\ref{fig:Dyn_grid}~(b) (the buses are chosen in terms of the increasing values of $K^{L^*}_{v,s}$). Recall that $K^{L^*}_{v,s}$ is the true value of the attack controller gain at which the system becomes unstable and $\widehat{K}^{L^*}_{v,s}$ is the prediction based on the sensitivity analysis.
We observe that bus~$19$ has the least value of $K^{L^*}_{v,s},$ and thus $v^* = 19.$ Moreover, the values of $K^{L^*}_{v,s}$ and  $\widehat{K}^{L^*}_{v,s}$ match closely. This result shows that the parametric sensitivity approach can accurately predict the critical vulnerable nodes of the system. 


\begin{figure}[!t]
\centering
\includegraphics[width=0.48\textwidth]{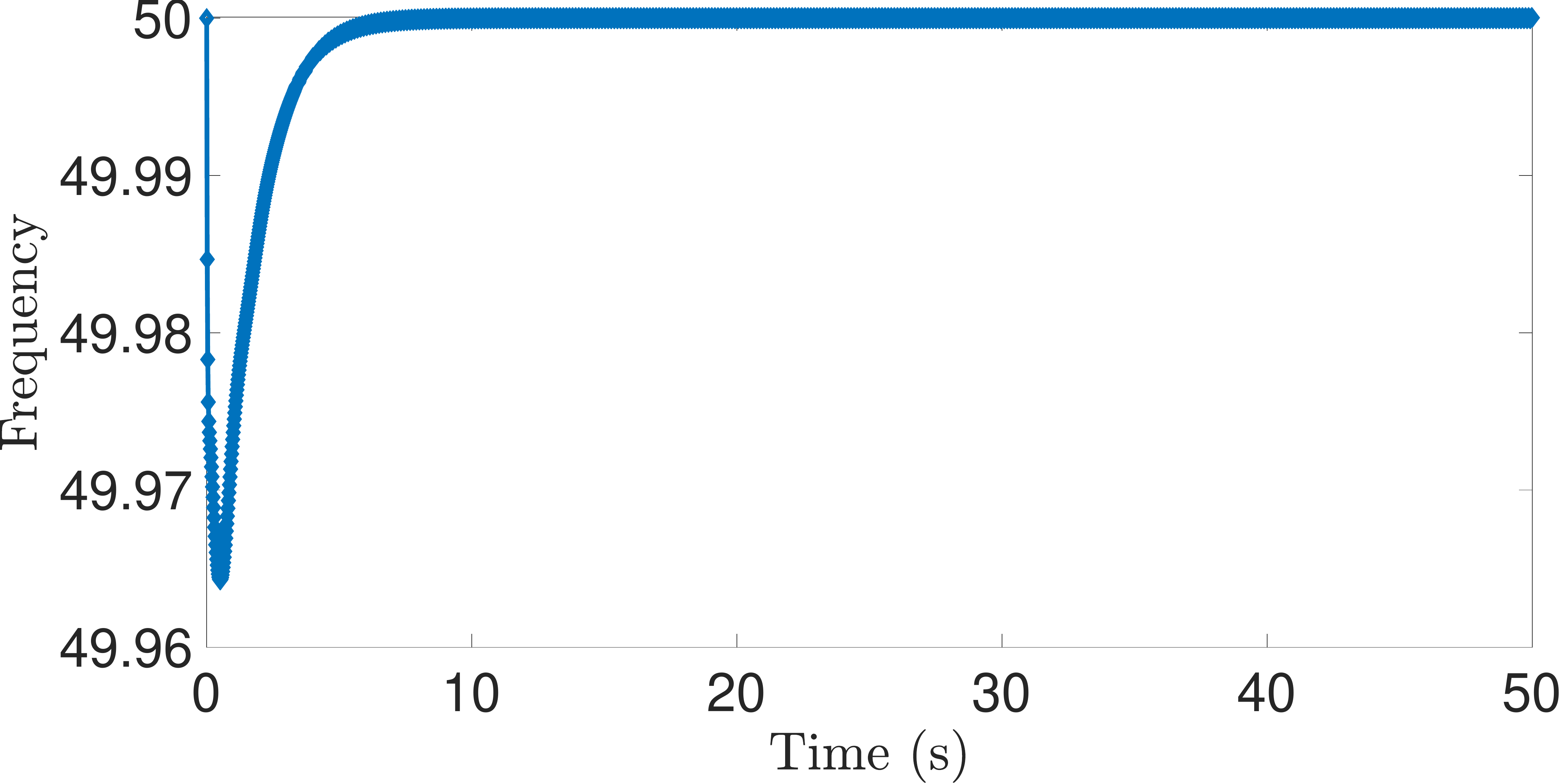}
\caption{Frequency dynamics of generator bus~30 under SLAA. $\epsilon_v = 2~\text{p.u.}$ with $v = 19$ for the IEEE-39 bus system.}
\label{fig:Dynamics_39bus_dynstat}
\vspace{-0.2 cm}
\end{figure}

To compare the impact of DLAAs and SLAAs,  
we compromise an identical amount of load of 2~p.u. as an SLAA (one-time step change) at victim bus~19 and plot of frequency dynamics in Fig.~\ref{fig:Dynamics_39bus_dynstat}. 
It can be observed that in contrast to DLAAs, SLAA only leads to a minor deviation and the frequency gets restored to the nominal value relatively quickly. Thus, DLAA is clearly advantageous for the attacker. 

\begin{figure}[!t]
\centering
\begin{subfigure}{0.45\textwidth}
\includegraphics[width=1\textwidth]{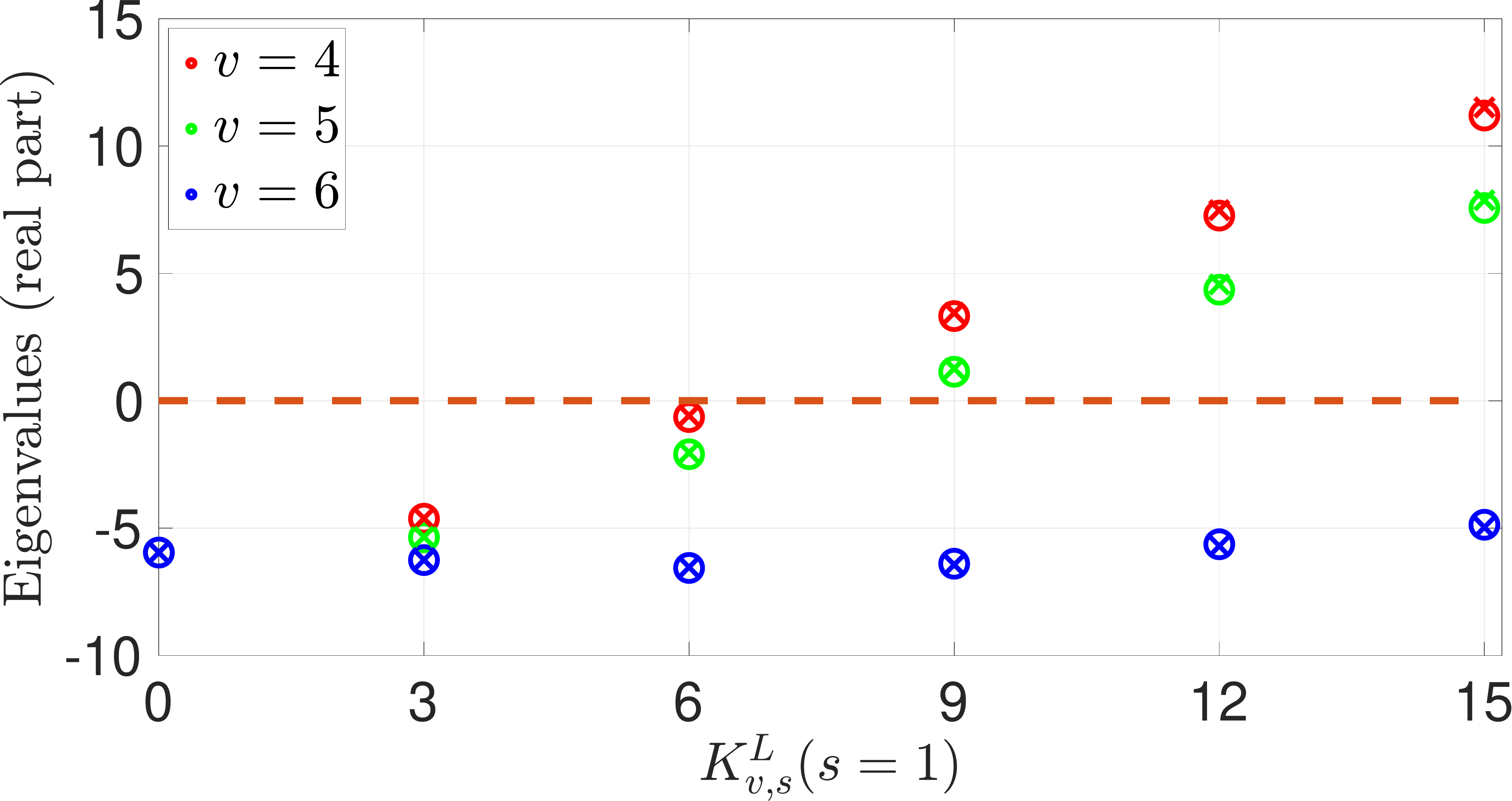}
\caption{}
\end{subfigure}
~
\begin{subfigure}{0.45\textwidth}
\includegraphics[width=1\textwidth]{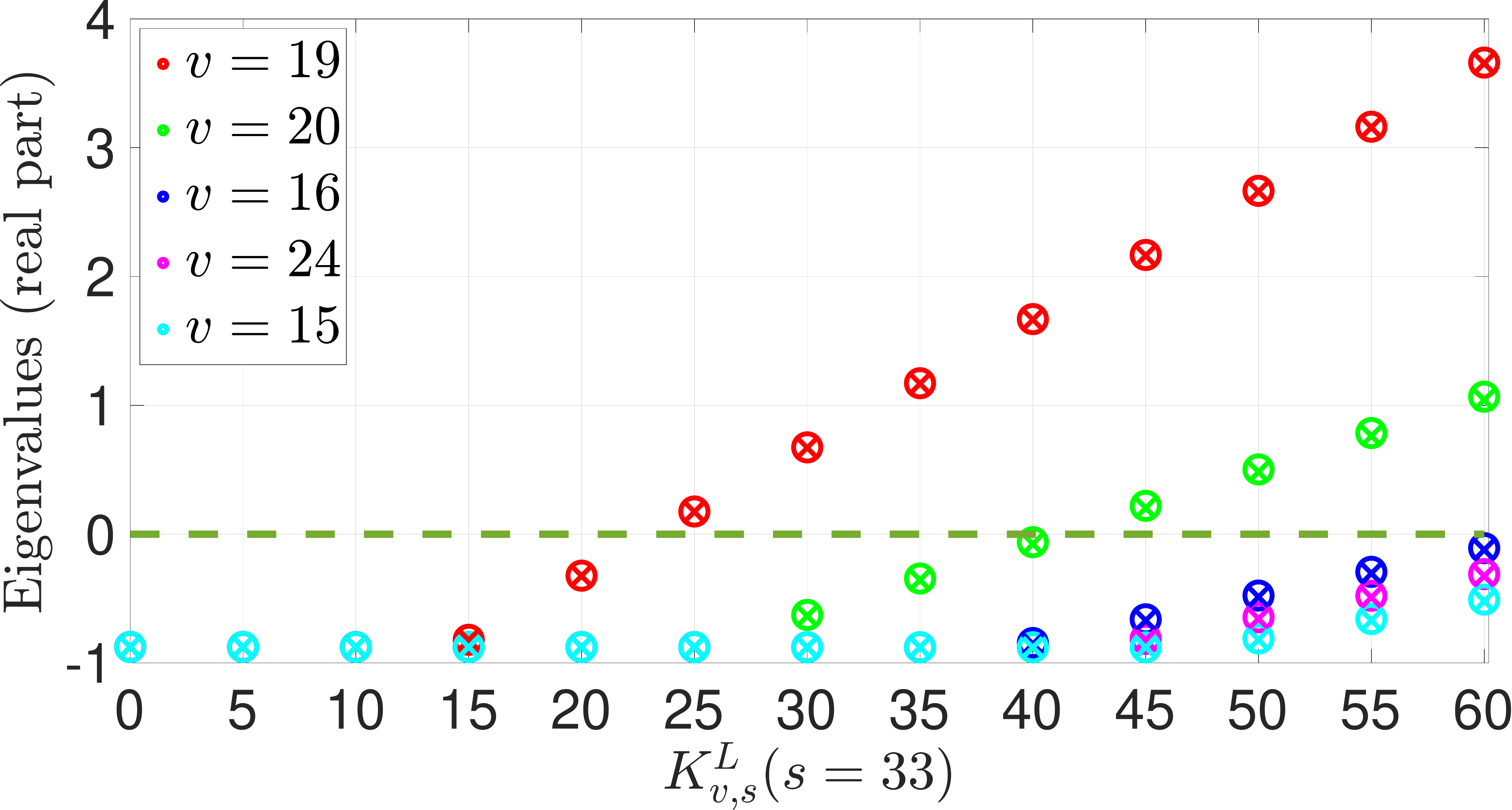}
\caption{}
\end{subfigure}
\caption{Real part of $\nu_1 (\Km^L) $ under single-point DLAAs for different values of $K^L_{v,s}.$ (a) IEEE-6 bus, $s = 1$ and $v = \{ 4 \},\{ 5 \}, \{ 6 \}  .$ (b) IEEE-39 bus, $s = 33$ and $v = \{ 19 \},\{ 20 \}, \{ 16 \} , \{ 24 \} , \{ 15 \}  $. Circles: ${\nu}_i ( \Km^L)$), Crosses: $\widehat{\nu}_i ( \Km^L)$.}
\label{fig:Eigvals_match}
\vspace{-0.2 cm}
\end{figure}

Further, we verify the accuracy of the sensitivity approach in approximating the true eigenvalues of the system under DLAAs. 
Without the loss of generality, assume that eigenvalues $\{ \nu_i (\Km^L) \}^{2N_g}_{i = 1} $ are sorted according to the decreasing value of their real parts. We plot the real part of $\nu_1 (\Km^L) $ (i.e., eigenvalue which has the maximum real part) and $\widehat{\nu}_1 ( (\Km^L))$ for IEEE-6 bus and IEEE-39 bus systems by varying $K^L_{v,s}$ in Fig.~\ref{fig:Eigvals_match}. For the IEEE-6 bus system, $s = 1$ and $v = \{ 4 \},\{ 5 \}, \{ 6 \}  .$ 
For the IEEE-39 bus system, $s = 33$ and $v = \{ 19 \},\{ 20 \}, \{ 16 \} , \{ 24 \} , \{ 15 \}  $ (we choose the $5$ victim buses that correspond to the $5$ least-effort DLAAs in this case). 
We observe that for both the bus systems, the two quantities match closely, thus validating the parametric sensitivity approach. Further, we also observe that the accuracy degrades slightly for large values of attack controller gains  $K^L_{v,s},$ which is expected since the sensitivity approach is a linear approximation. However, for a reasonable range of $K^L_{v,s},$ the approximation remains accurate. E.g., in Fig.~\ref{fig:Eigvals_match} (bottom figure), we observe a good match for $K^L_{v,s},$ values up to $50~$p.u., which corresponds to $p^L_v = 2 K^L_{v,s} \omega_{\max} = 2 \times 50 \times 2/50 = 4 \text{p.u.} = 400~\text{MWs}$ of compromised load (assuming base load of $100~$MWs). 

We enlist the parameter $\eta = (K^{L^*}_{v^*,s} - \widehat{K}^{L^*}_{v^*,s})/K^{L^*}_{v^*,s}$ for different IEEE bus systems in Table~\ref{tbl:Accuracy}. Recall that $\{v^*\} = \arg \min_{v \in \mathcal{V}}  \widehat{K}^{L^*}_{v,s},$ and thus, we are identifying the attack parameters that correspond to the least-effort destabilizing attacks for different bus systems.
We see that the sensitivity approach can closely approximate the value of attack controller gain at which the system becomes unstable. Further, we observe that the accuracy of the sensitivity approach is independent of the size of the bus system under consideration, but however, depends on the value of $K^L_{v,s}.$ This is evident from the values of $K^{L^*}_{v^*,s}$ noted in Table~\ref{tbl:Accuracy}, where we observe that the accuracy slightly degrades for higher values of $\widehat{K}^{L^*}_{v,s},$ which is consistent with the observation in Fig.~\ref{fig:Eigvals_match}~(b).


\begin{table}[!t]
 \begin{center}
 \caption{Value of $\eta = \big{|} (K^{L^*}_{v^*,s} - \widehat{K}^{L^*}_{v^*,s})/K^{L^*}_{v^*,s} \big{|}$ for different IEEE bus systems. Bus~1 is assumed to be sensor bus.}
 \label{tbl:Accuracy}
 \begin{tabular}{|c | c | c | c | c |c |} 
 \hline
 Bus system & Sensing bus & $\eta$ &  $K^{L^*}_{v^*,s}$ \\ [0.5ex] 
 \hline\hline
 IEEE 6-bus system & 1 & $0$ & $6.1$ \\ 
 \hline
 IEEE 14-bus system & 1 & $0.009$  &  $11$\\
 \hline
 IEEE 39-bus system & 30 & $0.0385$  & $249$\\
 \hline
 IEEE 39-bus system & 33 & $0.0043$  & $23.4$\\
 \hline
 IEEE 39-bus system & 37 & $0.0059$  & $58.3$ \\
  [1ex] 
 \hline
\end{tabular}
\end{center}
\vspace{-0.4 cm}
\end{table} 



\begin{figure}[!t]
	\centering
		\includegraphics[width=0.48\textwidth]{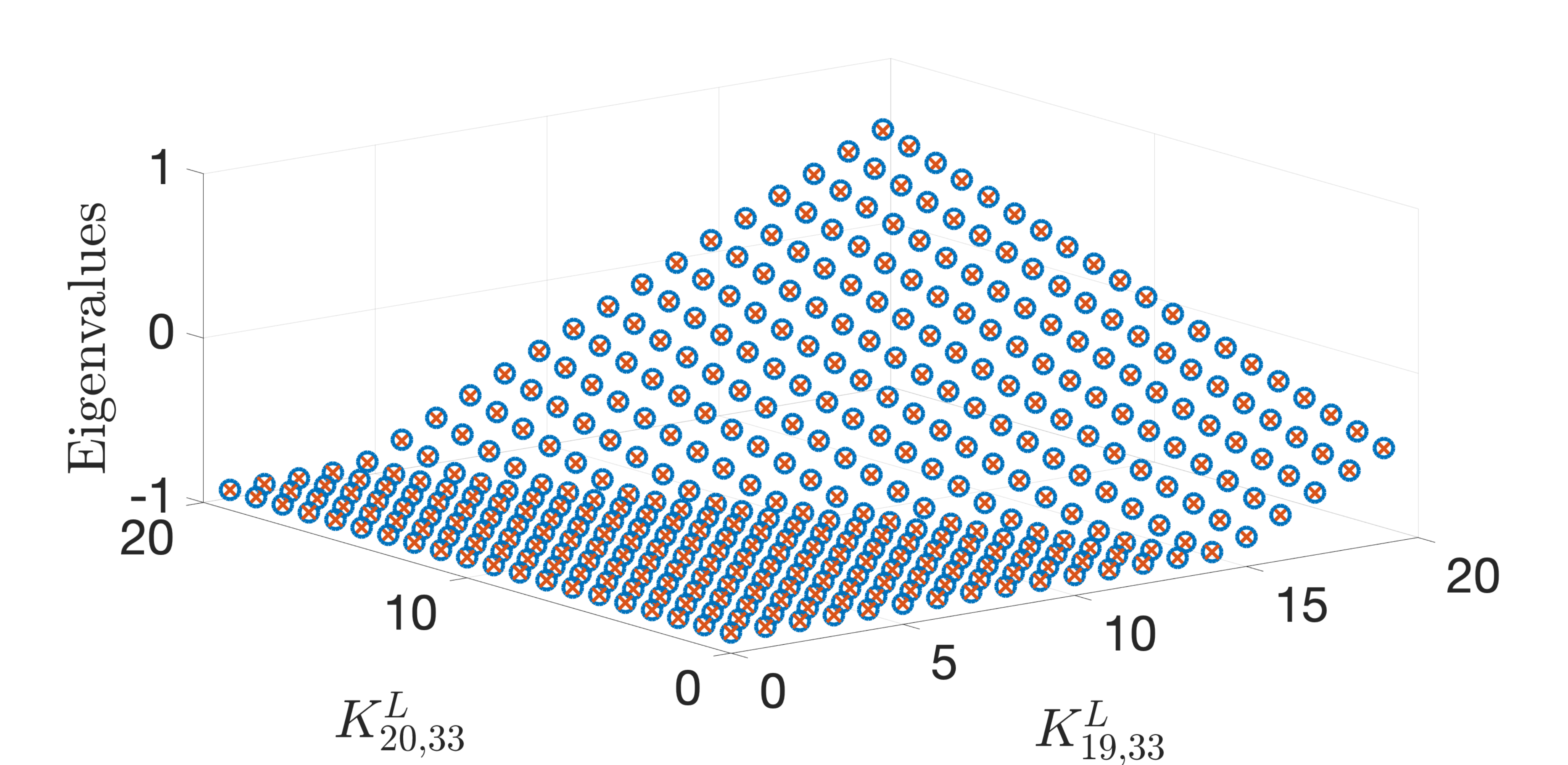}
	\caption{Real part of $\nu_1 (\Km^L) $ under multipoint DLAAs for different values of $K^L_{19,33}$ and $K^L_{20,33}$ for IEEE-39 Bus system. Circles: ${\nu}_1 ( \Km^L)$, Crosses: $ \widehat{\nu}_1 ( \Km^L)$. }
	\label{fig:freq_timecross}
	\vspace{-0.3cm}
\end{figure}

{\bf Multi-Point DLAAs:}
Next, we investigate multi-point DLAAs. We vary the attack controller gain values of two victim nodes simultaneously in the IEEE-39 bus system, namely buses 19 and 20 (note these two victim bus correspond to the locations of the least-effort load-altering attack). 
We plot the true eigenvalues of the system ${\nu}_1 ( \Km^L)$ and those predicted by the sensitivity approach $ \widehat{\nu}_1 ( \Km^L)$ in Fig.~\ref{fig:freq_timecross}. Once again, we observe a close match between the two, showing that the proposed approach is effective in approximating the true eigenvalues under multi-point DLAAs.  

\begin{figure}[!t]
\centering
\begin{subfigure}{0.45\textwidth}
\includegraphics[width=1\textwidth]{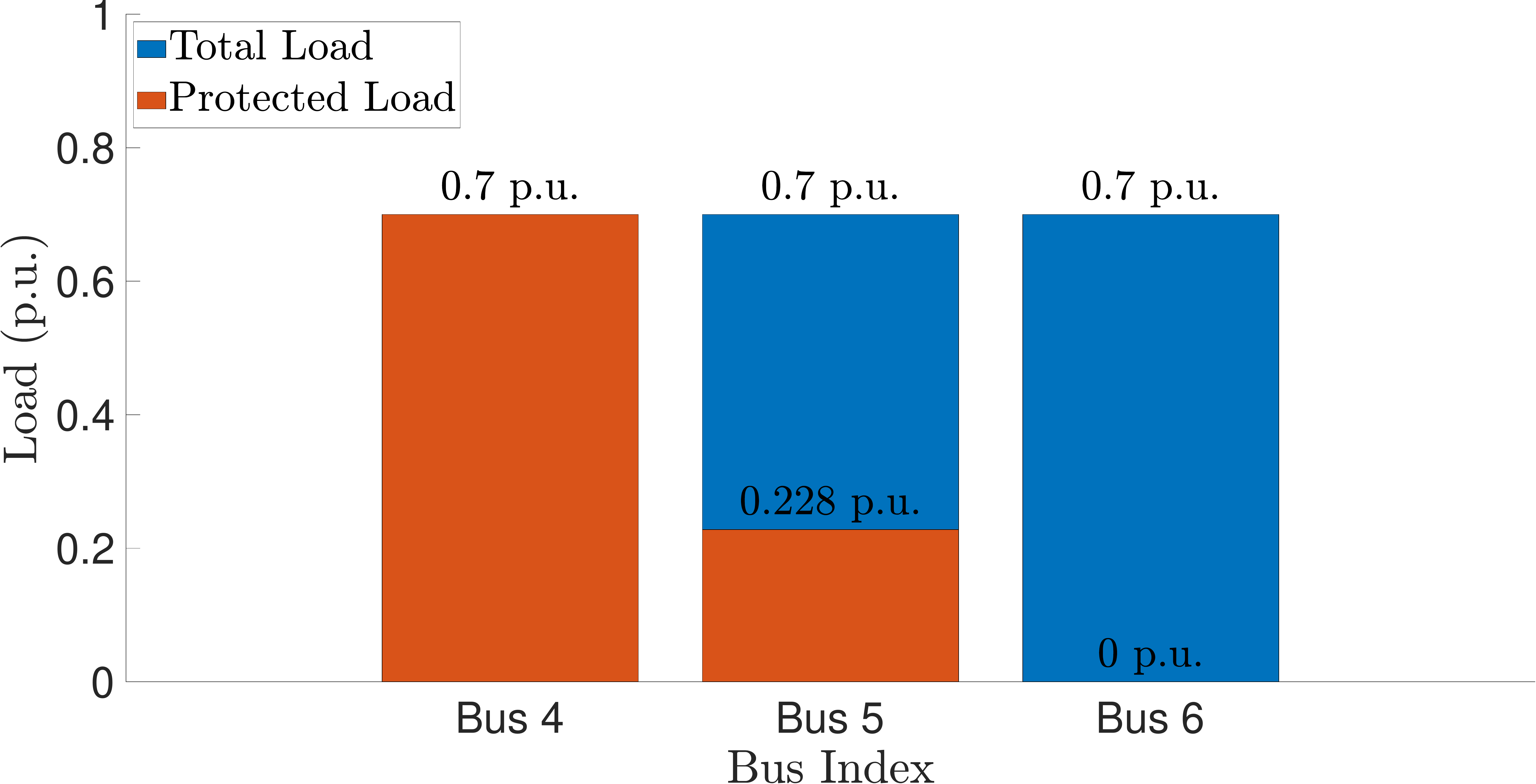}
\caption{}
\end{subfigure}
~
\begin{subfigure}{0.45\textwidth}
\includegraphics[width=1\textwidth]{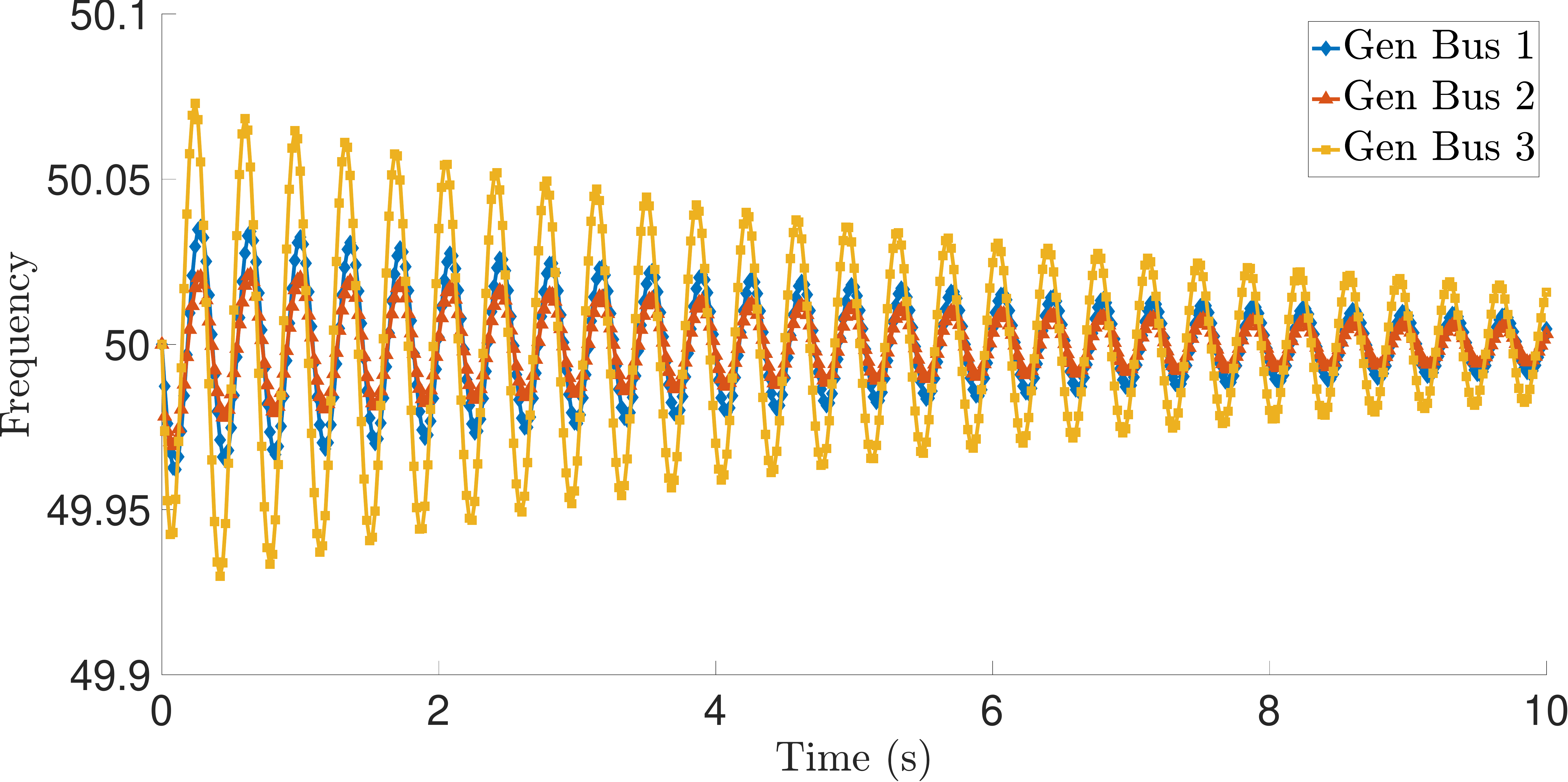}
\caption{}
\end{subfigure}
\caption{(a) Protected load to defend against DLAAs. (b) Dynamics under multi-point DLAAs with the unprotected load. Both plots consider the IEEE 6-bus system.}
\label{fig:protection}
\end{figure}

\begin{figure}[!t]
	\centering
		\includegraphics[width=0.48\textwidth]{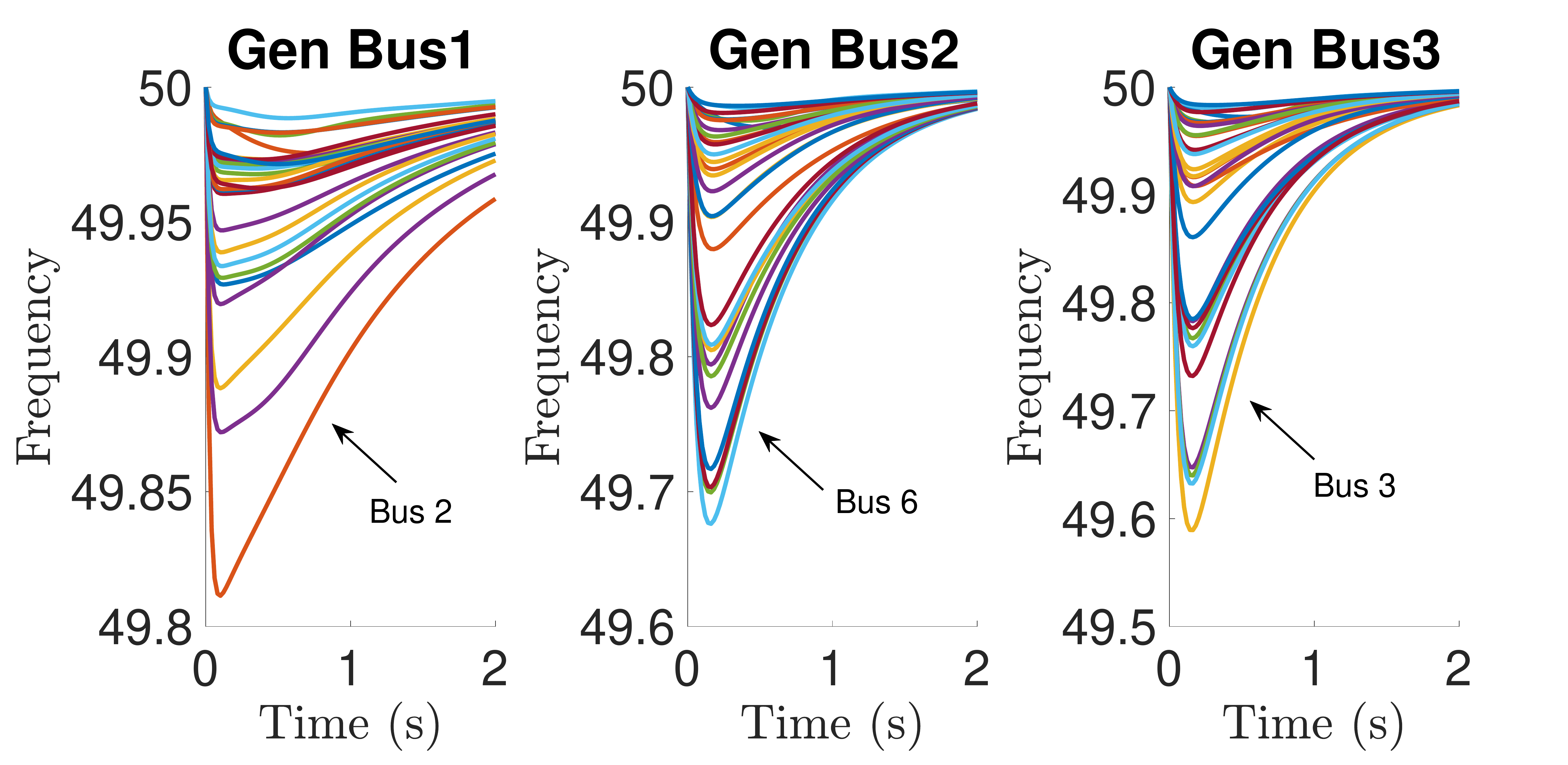}
	\caption{Function $f_{i,n}(t)$ for different generator and load buses in IEEE-39 bus system. Each curve corresponds to a victim bus. The victim bus corresponding to the least-effort SLAA is marked.}
	\label{fig:Perunit_fluctuations}
\end{figure}

We also implement the defense against DLAAs by solving the optimization problem \eqref{eqn:Defense_DLAA}. We plot the amount of load to be protected according to the solution of \eqref{eqn:Defense_DLAA} for the IEEE-6 bus system in Fig.~\ref{fig:protection} (a). To verify its correctness, we plot the frequency dynamics considering the maximum permissible values of attack feedback again, i.e., by setting ${K}^L_{v,s}= (P^{LV}_v - {P^{LP}}_v^*)/2 \omega^{\max}_s, $ where (${P^{LP}}^*_v$ is the solution of \eqref{eqn:Defense_DLAA}) in Fig.~\ref{fig:protection} (b). We observe that the oscillations are damping and will eventually go to $0$, thus verifying that the proposed defense can make the system resilient to DLAAs.

{\bf Static Load Altering Attacks:}
We also perform simulations for the results derived from SLAA in Section~\ref{sec:SLAA}. To this end, we plot the functions $f_{i,n}(t)$ for different generator and load buses considering IEEE-39 bus system in Fig.~\ref{fig:Perunit_fluctuations}. These functions represent the fluctuation of the frequency for per unit change in the load. For the ease of illustration, we only plot the curves corresponding to three generator buses, i.e., bus~29,30 and 31. The victim bus corresponding to the least-effort SLAA is marked in the figure.
Using the curves above and the result in \eqref{eqn:least_load}, a grid operator can  determine the minimum amount of load altering required to cause unsafe frequency fluctuation. 

{\bf Comparison with Non-linear Model:} We also compare the effectiveness of the proposed sensitivity approach in predicting the 
least-effort DLAA under a non-linear model of the power grid. 
To this end, we simulate a simplified version of the non-linear model given by
\begin{align*}
&  \dot{\delta}_i = \omega_i,  i \in \mathcal{N}_G \\
& M_i \dot{\omega}_i = - D_i \omega_i - K^P_i  \omega_i - K^I_i  \delta_i ,  \qquad \qquad  i \in \mathcal{N}_G \nonumber \\ &  \ - \sum_{j \in \mathcal{N}_G} B_{i,j} \sin (\delta_i - \delta_j)  - \sum_{j \in \mathcal{N}_L} B_{i,j} \sin (\delta_i - \theta_j), \\
D_i \dot{\theta}_i &=  \sum_{s \in \mathcal{S}} K^{LG}_{i,s} \omega_s  - P^{LS}_i  \qquad \qquad \qquad  \qquad i \in \mathcal{N}_L \nonumber \\ &  \ - \sum_{j \in \mathcal{N}_G} B_{i,j} \sin (\theta_i - \delta_j)  - \sum_{j \in \mathcal{N}_L} B_{i,j} \sin (\theta_i - \theta_j).
\end{align*}
By varying the attack controller gain values $K^{L}_{i,s} $ in the equations above, we find the minimum value of $K^{L^*}_{i,s} $ at which the system becomes unstable. We compare this with $\widehat{\nu}_i ( K^{L^*}_{v,s})$ obtained by the sensitivity approach (formulated based on the linear model). 
The results are listed in Table~\ref{tbl:Non_Linear} for different IEEE bus systems. We note that the sensitivity approach based on the linear model is able to predict the attack controller gain at which the non-linear system becomes unstable reasonably accurately. Thus, we believe that our analysis is a good initial step towards analyzing the system under more generalized system models.

\begin{table}[!t]
 \begin{center}
\caption{Comparison of the non-linear model with the sensitivity approach (based on the linear model) for different IEEE bus systems. }
\label{tbl:Non_Linear}
 \begin{tabular}{||c | p{1cm} | p{1.2 cm} | p{1.2cm}  |} 
 \hline
 Bus system & $\{v,s\}$ & ${K}^{L^*}_{v^*,s}$ (Non-linear) & $\widehat{\nu}_i ( K^{L^*}_{v,s})$ (Sensitivity)  \\ [0.5ex] 
 \hline\hline
  IEEE 6-bus system & $\{4,1\}$ & $6.7$ & $6.8$ \\ 
 \hline
 IEEE 14-bus system & $\{5,1\}$ & $11.3$  &  $11$\\
 \hline
  IEEE 39-bus system & $\{19,33\}$ & $25.2$  & $24.9$\\
 \hline
\end{tabular}
\end{center}
\vspace{-0.2 cm}
\end{table}

\section{Conclusions and Future Work}
\label{sec:Conc}
In this work, we have shown how results from second-order dynamical systems can be used to analyze IoT-based load altering attacks against power grids. Our results offer a low-complexity analytical approach to identify nodes corresponding to the least-effort destabilizing DLAAs and least-effort SLAAs that cause unsafe frequency excursions. Using these results, we also proposed defense against DLAAs and SLAAs. Our results show the analyses of DLAAs and SLAAs depend critically on the eigensolutions of the system and their sensitivity to changes in the attack parameters. Our analysis provides insights into how a grid operator can enhance the grid’s resilience to such attacks. To the best of our knowledge, this is the first work to apply concepts for second-order dynamical systems to analyze DLAAs and SLAAs.

There are several interesting future research directions. First, 
large-scale load-altering attacks might potentially result in major shifts in the dynamic/algebraic state of the power network, requiring analysis under generalized non-linear grid models rather than the linearized small-signal model used in this paper as well as prior works on this topic \cite{AminiLAA2018, AminiIdentification2019, AcharyaPHEV2020}. The preliminary simulation results presented in Section~\ref{sec:Results} suggest that the proposed sensitivity based approach could be a good starting point for this generalization. Moreover, this approach has been extended in the past to advanced systems involving general higher-order eigenvalue problems, see e.g., \cite{jp16}. Recent works \cite{li2014design,li2013study} also show that eigen-sensitivity analysis plays a significant role in the response analysis of general complex systems involving non-linear eigenproblems. Further research will be required in this direction to adapt the results for generalized models. Finally, analysis of the system that incorporates multiple control areas and potential safety mechanisms such as under frequency load shedding is important.

\balance
\bibliographystyle{IEEEtran}
\bibliography{IEEEabrv,bibliography,Eigenderivatives,StructuralMechanics,AdhikariPublications}

\begin{thebibliography}{10}
\providecommand{\url}[1]{#1}
\csname url@samestyle\endcsname
\providecommand{\newblock}{\relax}
\providecommand{\bibinfo}[2]{#2}
\providecommand{\BIBentrySTDinterwordspacing}{\spaceskip=0pt\relax}
\providecommand{\BIBentryALTinterwordstretchfactor}{4}
\providecommand{\BIBentryALTinterwordspacing}{\spaceskip=\fontdimen2\font plus
\BIBentryALTinterwordstretchfactor\fontdimen3\font minus
  \fontdimen4\font\relax}
\providecommand{\BIBforeignlanguage}[2]{{%
\expandafter\ifx\csname l@#1\endcsname\relax
\typeout{** WARNING: IEEEtran.bst: No hyphenation pattern has been}%
\typeout{** loaded for the language `#1'. Using the pattern for}%
\typeout{** the default language instead.}%
\else
\language=\csname l@#1\endcsname
\fi
#2}}
\providecommand{\BIBdecl}{\relax}
\BIBdecl

\bibitem{Bosch}
``{Bosch Home Connect},''
  \url{https://www.bosch-home.co.uk/bosch-innovations/homeconnect/}, Accessed:
  May 2020.

\bibitem{Fernandes2016HomeApp}
E.~{Fernandes}, J.~{Jung}, and A.~{Prakash}, ``Security analysis of emerging
  smart home applications,'' in \emph{Proc. IEEE Symposium on Security and
  Privacy (S\&P)}, May 2016, pp. 636--654.

\bibitem{maple2017security}
C.~Maple, ``Security and privacy in the internet of things,'' \emph{Journal of
  Cyber Policy}, vol.~2, no.~2, pp. 155--184, 2017.

\bibitem{Liu2009}
Y.~Liu, P.~Ning, and M.~K. Reiter, ``False data injection attacks against state
  estimation in electric power grids,'' in \emph{Proc. ACM Conference on
  Computer and Communications Security (CCS)}, 2009, pp. 21--32.

\bibitem{RenLoadRedis2011}
Y.~Yuan, Z.~Li, and K.~Ren, ``Modeling load redistribution attacks in power
  systems,'' \emph{IEEE Trans. Smart Grid}, vol.~2, no.~2, 2011.

\bibitem{LaksheEnergy2017}
S.~Lakshminarayana, T.~Z. Teng, D.~K.~Y. Yau, and R.~Tan, ``Optimal attack
  against cyber-physical control systems with reactive attack mitigation,'' in
  \emph{Proc. ACM International Conference on Future Energy Systems
  (e-Energy)}, 2017, pp. 179--190.

\bibitem{LakshDataDrive2020}
S.~{Lakshminarayana}, A.~{Kammoun}, M.~{Debbah}, and H.~V. {Poor},
  ``Data-driven false data injection attacks against power grids: {A} random
  matrix approach,'' \emph{IEEE Trans. Smart Grid}, 2020, doi:
  10.1109/TSG.2020.3011391.

\bibitem{LakshCCPA2019}
S.~Lakshminarayana, E.~V. {Belmega}, and H.~V. {Poor}, ``Moving-target defense
  for detecting coordinated cyber-physical attacks in power grids,'' in
  \emph{Proc. IEEE SmartGridComm}, Oct 2019, pp. 1--7.

\bibitem{HamedLAA2011}
A.~{Mohsenian-Rad} and A.~{Leon-Garcia}, ``Distributed internet-based load
  altering attacks against smart power grids,'' \emph{IEEE Trans. Smart Grid},
  vol.~2, no.~4, pp. 667--674, 2011.

\bibitem{Dabrowski2017}
A.~Dabrowski, J.~Ullrich, and E.~R. Weippl, ``Grid shock: {C}oordinated
  load-changing attacks on power grids: The non-smart power grid is vulnerable
  to cyber attacks as well,'' in \emph{Proc. ACSAC}, 2017, pp. 303--314.

\bibitem{Dvorking2017}
Y.~{Dvorkin} and S.~{Garg}, ``{IoT-enabled distributed cyber-attacks on
  transmission and distribution grids},'' in \emph{Proc. North American Power
  Symposium (NAPS)}, 2017, pp. 1--6.

\bibitem{Soltan2018}
S.~Soltan, P.~Mittal, and H.~V. Poor, ``{BlackIoT: IoT} botnet of high wattage
  devices can disrupt the power grid,'' in \emph{Proc. {USENIX} Security
  Symposium}, Baltimore, MD, Aug. 2018, pp. 15--32.

\bibitem{HuangUSENIX2019}
B.~Huang, A.~A. Cardenas, and R.~Baldick, ``Not everything is dark and gloomy:
  Power grid protections against iot demand attacks,'' in \emph{Proc. {USENIX}
  Security Symposium}, Aug. 2019, pp. 1115--1132.

\bibitem{AminiLAA2018}
S.~{Amini}, F.~{Pasqualetti}, and H.~{Mohsenian-Rad}, ``Dynamic load altering
  attacks against power system stability: Attack models and protection
  schemes,'' \emph{IEEE Transactions on Smart Grid}, vol.~9, no.~4, pp.
  2862--2872, July 2018.

\bibitem{AminiIdentification2019}
S.~{Amini}, F.~{Pasqualetti}, M.~{Abbaszadeh}, and H.~{Mohsenian-Rad},
  ``Hierarchical location identification of destabilizing faults and attacks in
  power systems: A frequency-domain approach,'' \emph{IEEE Transactions on
  Smart Grid}, vol.~10, no.~2, pp. 2036--2045, 2019.

\bibitem{FreqMes}
``{Measuring devices for frequency measurement},''
  \url{https://www.mainsfrequency.com/meter.htm}, Accessed: Jan 2021.

\bibitem{ZhaoFreq2013}
C.~{Zhao}, U.~{Topcu}, and S.~H. {Low}, ``Optimal load control via frequency
  measurement and neighborhood area communication,'' \emph{IEEE Transactions on
  Power Systems}, vol.~28, no.~4, pp. 3576--3587, 2013.

\bibitem{DLAAStorage2020}
R.~{Germanà}, A.~{Giuseppi}, and A.~{Di Giorgio}, ``Ensuring the stability of
  power systems against dynamic load altering attacks: A robust control scheme
  using energy storage systems,'' in \emph{Proc. European Control Conference
  (ECC)}, 2020, pp. 1330--1335.

\bibitem{AcharyaPHEV2020}
S.~{Acharya}, Y.~{Dvorkin}, and R.~{Karri}, ``Public plug-in electric vehicles
  + grid data: {I}s a new cyberattack vector viable?'' \emph{IEEE Transactions
  on Smart Grid}, vol.~11, no.~6, pp. 5099--5113, 2020.

\bibitem{book1b}
\BIBentryALTinterwordspacing
S.~Adhikari, \emph{Structural Dynamic Analysis with Generalized Damping Models:
  Identification}.\hskip 1em plus 0.5em minus 0.4em\relax UK: Wiley ISTE, 2013,
  (272 pages). [Online]. Available:
  \url{http://eu.wiley.com/WileyCDA/WileyTitle/productCd-184821670X.html}
\BIBentrySTDinterwordspacing

\bibitem{mei97}
L.~Meirovitch, \emph{Principles and Techniques of Vibrations}.\hskip 1em plus
  0.5em minus 0.4em\relax New Jersey: Prentice-Hall International, Inc., 1997.

\bibitem{Smed1993}
T.~{Smed}, ``Feasible eigenvalue sensitivity for large power systems,''
  \emph{IEEE Trans. Power Syst.}, vol.~8, no.~2, pp. 555--563, 1993.

\bibitem{NamEigSens2000}
{Hae-Kon Nam}, {Yong-Ku Kim}, {Kwan-Shik Shim}, and K.~Y. {Lee}, ``A new
  eigen-sensitivity theory of augmented matrix and its applications to power
  system stability analysis,'' \emph{IEEE Trans. Power Syst.}, vol.~15, no.~1,
  pp. 363--369, 2000.

\bibitem{kundur1994}
P.~Kundur, N.~Balu, and M.~Lauby, \emph{Power System Stability and
  Control}.\hskip 1em plus 0.5em minus 0.4em\relax McGraw-Hill Education, 1994.

\bibitem{ray1877}
J.~W. Rayleigh, \emph{Theory of Sound (two volumes)}, 1945th~ed.\hskip 1em plus
  0.5em minus 0.4em\relax New York: Dover Publications, 1877.

\bibitem{jp8}
S.~Adhikari, ``On symmetrizable systems of second kind,'' \emph{Transactions of
  ASME, Journal of Applied Mechanics}, vol.~67, no.~4, pp. 797--802, December
  2000.

\bibitem{li2014hybrid}
L.~Li, Y.~Hu, X.~Wang, and L.~L{\"u}, ``A hybrid expansion method for frequency
  response functions of non-proportionally damped systems,'' \emph{Mechanical
  Systems and Signal Processing}, vol.~42, no. 1-2, pp. 31--41, 2014.

\bibitem{mei80}
L.~Meirovitch, \emph{Computational Methods in Structural Dynamics}.\hskip 1em
  plus 0.5em minus 0.4em\relax Netherlands: Sijthoff \& Noordohoff, 1980.

\bibitem{jp7}
S.~Adhikari, ``Modal analysis of linear asymmetric non-conservative systems,''
  \emph{ASCE Journal of Engineering Mechanics}, vol. 125, no.~12, pp.
  1372--1379, December 1999.

\bibitem{jp14}
S.~Adhikari and M.~I. Friswell, ``Eigenderivative analysis of asymmetric
  non-conservative systems,'' \emph{International Journal for Numerical Methods
  in Engineering}, vol.~51, no.~6, pp. 709--733, June 2001.

\bibitem{TechReport}
S.~Lakshminarayana, S.~Adhikari, and C.~Maple, ``Techical report,''
  \url{https://bit.ly/3ryt0zA}, Accessed: Mar 2021.

\bibitem{jp16}
S.~Adhikari, ``Derivative of eigensolutions of non-viscously damped linear
  systems,'' \emph{AIAA Journal}, vol.~40, no.~10, pp. 2061--2069, October
  2002.

\bibitem{li2014design}
L.~Li, Y.~Hu, and X.~Wang, ``Design sensitivity and hessian matrix of
  generalized eigenproblems,'' \emph{Mechanical Systems and Signal Processing},
  vol.~43, no. 1-2, pp. 272--294, 2014.

\bibitem{li2013study}
------, ``A study on design sensitivity analysis for general nonlinear
  eigenproblems,'' \emph{Mechanical Systems and Signal Processing}, vol.~34,
  no. 1-2, pp. 88--105, 2013.

\end{thebibliography}

\end{document}